\documentclass[aps,twocolumn,pre,showpacs,superscriptaddress]{revtex4-1}%
\usepackage{graphicx}
\usepackage{amssymb}
\usepackage{bm}

\usepackage{epsfig}

\usepackage{amsmath}
\usepackage{bbm} 
\usepackage{makeidx} 
\usepackage{wrapfig}
\usepackage{float}
\usepackage{pstricks,pst-grad,pst-plot}
\usepackage{pst-node}
\usepackage{graphicx}
\usepackage{dcolumn}
\usepackage{bm}
\usepackage{amsmath}
\usepackage{amssymb}
\usepackage{epsfig}
\usepackage{verbatim}
\usepackage{pst-grad,pst-node,pst-tree,pst-plot}
\usepackage{pstricks}
\usepackage{natbib}
\usepackage{ifthen}
\usepackage{comment}

\begin{document}

\title{Mechanisms of evolution of avalanches in regular graphs}

\author{Thomas~P.~Handford}
\affiliation{Department of Chemistry, University of Cambridge, Cambridge, UK}
\author{Francisco~J.~P{\'e}rez-Reche}
\affiliation{Institute for Complex Systems and Mathematical Biology, SUPA, King's College, University of Aberdeen, Aberdeen, UK }
\author{Sergei~N.~Taraskin}
\affiliation{St. Catharine's College and Department of Chemistry,
University of Cambridge, Cambridge, UK}

\begin{abstract}
A mapping of avalanches occurring in the zero-temperature random-field Ising model (zt-RFIM) to life-periods of a population experiencing immigration is established.
Such a mapping allows the microscopic criteria for occurrence of an infinite avalanche in a $q$-regular graph to be determined. 
A key factor for an avalanche of spin flips to become infinite is that it interacts in an optimal way with previously flipped spins.
Based on these criteria, we explain why an infinite avalanche can occur in $q$-regular graphs only for $q>3$, and suggest that this criterion might be relevant for other systems.
The generating function techniques developed for branching processes are applied to obtain analytical expressions for the duration, pulse-shapes and power spectrum of the avalanches. 
The results show that only very long avalanches exhibit a significant degree of universality.
\end{abstract}


\maketitle

\section{Introduction}

Systems exhibiting avalanche response when driven externally are ubiquitous and have been widely studied in recent decades~\cite{Sethna2001}. 
Examples of such phenomena include magnetisation reversal~\cite{Durin_review2004}, collapse of overloaded materials~\cite{Pradhan2001,Pradhan2009}, earthquakes~\cite{Carlson1989,Baro_PRL2013} 
or collective opinion shifts~\cite{Michard-Bouchaud_EPJB2005_RFIM-OpinionDynamics}. Broadly speaking, this kind of systems can either pop (all the avalanches are of similar size), snap (there is at least one characteristic large event) or crackle (avalanches have a broad range of sizes). 
The competition between spatial interactions promoting snap behaviour and heterogeneity that favour popping has been proposed as a key factor dictating whether a system will snap, pop or crackle~\cite{Sethna2001}. 
However, the mechanisms which determine the nature of such avalanches  are still poorly understood and this affects the ability to make predictions. 
For instance, it is often a challenging task to predict the occurrence of large avalanches (e.g. a devastating earthquake or an abrupt collective change of opinion) by observing the evolution of systems over a short period of time.

In this paper, we investigate the link between microscopic dynamics and the type of global avalanche response within the framework of the zero-temperature random-field Ising model (zt-RFIM)~\cite{Sethna1993,Sethna2001}. 
This prototype model, originally proposed for disordered magnetic materials, provides a unified description of the three common avalanche types in various systems.
We demonstrate that the time-dependent behaviour of avalanches in the zt-RFIM defined on locally tree-like topologies (e.g. a $q$-regular graph, where $q$ is the coordination number) is described by a branching process (BP) which is a well-established model for population dynamics~\cite{Jagers2005}. 
The mapping provides the criteria for the appearance of an infinite avalanche in the system. 
In particular, we give an explanation for the long-standing question~\cite{Sabhapandit2000,SabhapanditPRL} of why infinite avalanches can occur in a Bethe lattice or a $q$-regular graph with $q>3$, but do not occur for $q\le 3$. 
In addition, the proposed framework leads to exact solutions for experimentally important quantities such as the distribution of avalanche durations, their mean pulse shape and mean power spectrum.

This paper is structured as follows. 
The  zt-RFIM is briefly described in Sec.~\ref{sec:Model}. 
In Sec.~\ref{sec:Dynamics}, the mapping between the zt-RFIM and BP for a $q$-regular graph is established. 
A necessary condition for an infinite avalanche to occur is then presented in Sec.~\ref{sec:Condition}. 
In Sec.~\ref{sec:TimeDep}, the derivation of the time-dependent properties of avalanches based on the generating function technique is given and conclusions follow in Sec.~\ref{sec:Conclusions}.

\section{The Model}\label{sec:Model}

The zt-RFIM considers a set of $N$ spin variables $s_i=\pm 1$ placed on the nodes of a network with links representing interaction between pairs of spins. 
The system is described by the following Hamiltonian,
\begin{equation}
{\cal H} = -J\sum_{\langle i,j\rangle}^Ns_is_j-H(t)\sum_{i=1}^Ns_i-\sum_{i=1}^N{h_is_i}~,
\label{eq:Hamiltonian}
\end{equation}
where the first summation is taken over all pairs of interacting sites, $\langle i,j\rangle$ and interaction strength $J=1$.
Heterogeneity is introduced by quenched and independent local random fields, $h_i$, distributed with the probability density function (p.d.f.) $\rho(h_i)$, which is continuous and non-zero everywhere and characterised by a zero mean and typical variance $\Delta^2$.
The system is driven by the external field, $H(t)$, which is assumed to increase monotonically and adiabatically from $-\infty$ (when all spins are in the down-state, $s_i=-1$) to $+\infty$. 
Spins flip to the up-state ($s_i=+1$) at a rate $\Gamma$ according to single spin-flip dynamics~\cite{Sethna1993}, i.e. they align with their local fields, 
$f_i(t)=H(t)+h_i+J(2n_i(t)-q)$,
where $n_i(t)=\sum_{j/i}\left(s_j+1\right)/2$ is the number of spins $j$ which neighbour spin $i$ and are in the up-state and $q$ is the coordination number of node $i$ which is the same for all nodes in $q$-regular graph. 

With this dynamics, the system exhibits intermittent response with popping being observed for any network topology at large disorder. 
An infinite avalanche occurs at low disorder for certain network topologies:
a complete graph (mean-field model~\cite{Sethna1993}), a Bethe lattice or a $q$-regular graph with coordination number $q>3$~\cite{Dhar1997}, and on hypercubic lattices in dimensions $d\ge 3$~\cite{Perkovic1999,Spasojevi2011,PerezReche2003,PerezReche2004RFIMField}. 
For these topologies, pop and snap regimes are separated by a disorder-induced continuous phase transition located at a topology-dependent critical degree of disorder, $\Delta=\Delta_c$, where crackling occurs. 
In contrast, no disorder-induced transition is observed for $q$-regular graphs with $q<3$ which exhibit popping for any non-zero disorder~\cite{Dhar1997}.

\section{Transformation dynamics of the zt-RFIM and mapping to a BP}\label{sec:Dynamics}

\begin{figure}
\includegraphics{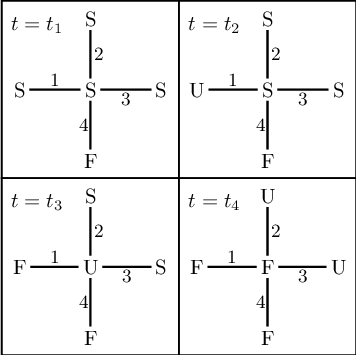}
\caption{Example of an avalanche on part of a 4-regular graph. 
At time $t_1$, the configuration consists of only S (stable) and F (flipped) spins. 
As the external field increases, a spin becomes unstable (U) at time $t_2>t_1$. 
This represents an immigration event for the population of U spins. 
At time $t_3>t_2$, the U spin flips causing the central spin, $i$, to undergo a $S \to U$ transition. 
The central spin, $i$, had one flipped neighbour (along link 4) prior to time $t_3$, i.e. $n_i(t<t_3)=1$, and it became unstable, so that $\zeta_i$ is equal to the number of neighbouring spins in state S, i.e. $\zeta_i=2$. 
In contrast, $\zeta_i$ would have been zero if the central spin had remained stable. 
At time $t_4$, the central spin flips and this leads to $\xi_i=2$ new U offspring.
\label{fig:avalanche}
}
\end{figure}

In order to establish the time-dependent behaviour of the zt-RFIM and map it to a BP, it is convenient to introduce three states for spins: stable (S), unstable (U) and flipped (F). 
As the external field is driven from $-\infty$ to $\infty$, each spin passes consecutively through these states,  S$\to$U$\to$F (see an example in Fig.~\ref{fig:avalanche}). 
In the initial (stable, S) state, the local field for spin $i$ is negative, $f_i\le 0$, and thus it is in the down-state. 
At some time, the local field at spin $i$ becomes positive ($f_i>0$) and the spin changes to state U meaning it becomes unstable but is still in the down-state.
Finally, at a later time, spin $i$ stochastically flips and moves into the flipped state F characterised by $f_i>0$ and $s_i=+1$.
The first transition, S$\to$U, is mediated by the local field.
As such it can be caused by one of two mechanisms: 
(i) field-induced, i.e. due to the increase in external field $H(t)$ or 
(ii) flip-induced, caused by the flip of a neighbouring spin and associated increase in $n_i(t)$.
The second transition, U$\to$F, occurs as a Poisson process at rate $\Gamma$.
As a consequence of the second transition, a certain 
number $\xi_i$ of the neighbours of spin $i$ move from state S$\to$U. 

These transformations occurring in the zt-RFIM can
be mapped to a BP 
describing population dynamics
in which two types of event can take place: immigration of new individuals and reproduction followed by immediate death~\cite{Jagers2005}.
Indeed, the spins in state U form a population of individuals that are created spontaneously (immigration) by the field-induced mechanism and can produce new U spins (reproduction) when passing to the state F (death) according to the flip-induced mechanism. 
In the example shown in Fig.~\ref{fig:avalanche}, an immigration event occurs at time $t_2$ followed by the reproduction events $U \xrightarrow{t_3} U \xrightarrow{t_4} 2U$. 

In the BP describing the zt-RFIM, both immigration and reproduction are random events.
Consequently, the number of offspring produced by spin $i$ during reproduction, $\xi_i$, is a random number, and the process which determines it can be divided into the following two steps: 
first, the spin $i$ must flip, increasing the local field at $\zeta_i$ neighbouring stable spins, and second, some number, $\xi_i$, of these stable spins can become unstable if their local field becomes positive. 
Therefore, it is possible to define two forms of expectation value, $\mathbb{E}[X_i]$, which takes the mean of some arbitrary quantity $X_i$ measured at the flipping spins and $\mathbb{E}^{\text{F}}[X_j]$, which takes the mean of the quantity $X_j$ measured at the stable neighbours of the flipping spins. 
It can be shown for a q-regular graph that $\mathbb{E}[\xi_i]=\mathbb{E}^{\text{F}}[\zeta_j]$ (if we define $\zeta_j=0$ when spin $j$ does not flip, see App.~\ref{sec:DerivExi}). 

It is a known property of a BP that
if the expectation value $\mathbb{E}[\xi_i]< 1$, then the population size will be non-zero only for certain ``life-periods'' of short duration~\cite{Jagers2005}.
At the end of each life-period the population becomes extinct, with the next life-period being started by an immigration event.
Such life-periods correspond to popping in the zt-RFIM.
If $\mathbb{E}[\xi_i]>1$, the population size can grow exponentially,
analogous to an infinite avalanche of spin flips in the zt-RFIM, causing the system to snap.

The mapping between zt-RFIM and BP described above can be quantified by the following values:
(i) the rate, $R^{\text{I}}$, of spontaneous transitions S$\to$U in field-induced changes, i.e. the rate of immigration in the BP, and the rate, $\Gamma$, of transitions U$\to$F in flip-induced changes, i.e. the rate of reproduction and death; 
(ii) the probability, $P^{\text{SU}}$, that a stable spin becomes unstable, S$\to$U, when its neighbour changes from U to F (a flip-induced change describing reproduction); 
(iii) the probability distributions, $P^{\text{I}}(\zeta_i)$ and $P^{\text{R}}(\zeta_i)$, of the numbers of neighbours, $\zeta_i$, of site $i$ which are in state S immediately after site $i$ moves from S to U according to either a field- or flip-induced change, respectively and 
(iv) the probability distributions $P^{\text{I}}(\xi_i)$ and $P^{\text{R}}(\xi_i)$ of the numbers of offspring of an individual produced in a
field- and flip-induced change, respectively. 
The expectation value of $\xi_i$ can be evaluated either by averaging over immigrated individuals giving $\mathbb{E}[\xi^{\text{I}}_i]=\sum_{\xi^{\text{I}}_i=0}^q\xi^{\text{I}}_iP^{\text{I}}(\xi_i^{\text{I}})$, or over reproduced individuals resulting in 
$\mathbb{E}[\xi^{\text{R}}_i]=\sum_{\xi_i=0}^q\xi^{\text{R}}_iP^{\text{R}}(\xi^{\text{R}}_i)$, which are, in general, different from each other (the superscript R is suppressed in the rest of the paper). 

All these probabilities and p.d.f.'s 
can be calculated by extending the derivation of the mean magnetisation for a $q$-regular graph presented in Ref.~\cite{Dhar1997}.
Specifically, the results of that paper are that any spin in the system will be in state F with probability,
\begin{equation}
P=F_0(P^*,H(t))~,
\label{eq:P}
\end{equation}
where $H(t)$ is assumed to vary adiabatically slowly. For increasing $H(t)$, $P^*$ in Eq.~\eqref{eq:P} is the smallest solution to the self-consistent equation, 
\begin{equation}
P^*=F_1(P^*,H(t))~.
\label{eq:SelfConsistent}
\end{equation}
Here, the functions $F_m(P^*,H(t))$ ($m=0,1, \dots, q$),
\begin{equation}
F_m(P^*,H(t))=\sum_{n=0}^{q-m}{\binom{q-m}{n}}\left(P^*\right)^n\left(1-P^*\right)^{q-m-n}p_n~,\label{eq:Fm}
\end{equation}
are defined in terms of the probabilities, 
\begin{equation}
p_n=\int\limits_{-J(2n-q)-H}^{\infty}\rho(h)\text{d}h~,\label{eq:pn}
\end{equation}
that the local field at a spin is positive when it has $n$ flipped neighbours. 
The p.d.f. $\rho(h)$ is assumed to be positive for any finite $h$ and characterised by a continuous cumulative function.
In this case, $p_n$ are continuous functions of $H$ such that $0<p_0<\ldots<p_q<1$, implying through Eq.~\eqref{eq:SelfConsistent} that $0<P^*<1$.
The special case of a rectangular distribution, when $\rho(h)$ can be zero, has been analysed in Ref.~\cite{Sabhapandit2000}, where it was shown that the discontinuous behaviour in magnetisation associated with an infinite avalanche occurs for any regular graph, including a linear chain.

Extending the analysis of Ref.~\cite{Dhar1997} we find the rate of field-induced changes in the following form, 
\begin{equation}
R^{\text{I}}=N\left(\frac{\text{d}H}{\text{d}t}\right)\left(\frac{\partial F_0(P^*,H(t))}{\partial H}\right)~.\label{eq:RI}
\end{equation}
Similarly, the probability, $P^{\text{SU}}$, that a spin becomes unstable when one of its neighbours flips is given by,
\begin{equation}
P^{\text{SU}}=q^{-1}(1-P^*)^{-1}\left(\frac{\partial F_0(P^*,H(t))}{\partial P^*}\right)~.\label{eq:PSU}
\end{equation}
Using arguments similar to those presented in Ref.~\cite{Dhar1997}, it can be shown that the number of stable neighbours, $\zeta_i$, of a spin $i$, which undergoes a field-induced transition S$\to$U, is a random variable distributed according to,
\begin{eqnarray}
P^I(\zeta_i)&=&{\binom{q}{\zeta_i}}\left(P^*\right)^{q-\zeta_i}\left(1-P^*\right)^{\zeta_i}\nonumber\\&\times&\frac{\partial p_{q-\zeta_i}}{\partial H}\left(\frac{\partial F_0(P^*,H(t))}{\partial H}\right)^{-1}
\end{eqnarray}
and the number of stable neighbours of a spin undergoing a flip-induced transition S$\to$U is distributed according to,
\begin{eqnarray}
P^R(\zeta_i)&=&q{\binom{q-1}{\zeta_i}}\left(P^*\right)^{q-1-\zeta_i}\left(1-P^*\right)^{\zeta_i}\nonumber\\&\times&(p_{q-\zeta_i}-p_{q-1-\zeta_i})\left(\frac{\partial F_0(P^*,H(t))}{\partial P^*}\right)^{-1}~.\label{eq:zetaR}
\end{eqnarray}
In order to calculate the number $\xi_i$ of neighbours of a spin $i$ which become unstable when spin $i$ flips, we utilise the assumption that the external field $H(t)$ changes adiabatically slowly, meaning that probability for any neighbour of spin $i$ to change state
during the interval of time when 
spin $i$ is unstable tends to zero 
(except the special case when 
the infinite avalanche is occurring).
This means that when spin $i$ flips, each of the $\zeta_i$  
neighbours of spin $i$ which were stable when spin $i$ underwent the transition S$\to$U are still stable, and will become unstable as a result of the flip of spin $i$ independently with probability $P^{\text{SU}}$.
This results in the following distributions of numbers of offspring,
\begin{equation}
P^{\text{I}}(\xi^{\text{I}}_i)=\sum_{\zeta_i=\xi^{\text{I}}_i}^q{\binom{\zeta_i}{\xi^{\text{I}}_i}}\left(P^{\text{SU}}\right)^{\xi_i^{\text{I}}}\left(1-P^{\text{SU}}\right)^{\zeta_i-\xi_i^{\text{I}}}P^I(\zeta_i)~,\label{eq:ImmxiDist}
\end{equation}
and
\begin{equation}
P^{\text{R}}(\xi_i)=\sum_{\zeta_i=\xi_i}^q{\binom{\zeta_i}{\xi_i}}\left(P^{\text{SU}}\right)^{\xi_i}\left(1-P^{\text{SU}}\right)^{\zeta_i-\xi_i}P^R(\zeta_i)~.\label{eq:RepxiDist}
\end{equation}
The formulae given by Eqs.~\eqref{eq:RI}-\eqref{eq:RepxiDist} link the zt-RFIM to a BP describing the evolution of a population with immigration and reproduction followed by immediate death.

\section{Conditions for an Infinite avalanche}\label{sec:Condition}

As known from population dynamics, an infinite avalanche can occur only when the mean number of offspring $\mathbb{E}[\xi_i]>1$. 
However, applying this criterion to the zt-RFIM is challenging because the value of $\mathbb{E}[\xi_i]$ is a function of the time-dependent external field, $H(t)$, 
and is restricted by the time-dependent availability of S spins in the neighbourhood of U spins. 
This contrasts our model with BPs proposed for the description of avalanches in critical stationary states with time-independent 
$\mathbb{E}[\xi_i]$~\cite{Alstrom_PRA1988,Kinouchi_PRE1999,Goh_PRL2003}
For the zt-RFIM,
$\mathbb{E}[\xi_i]=0$ at the initial moment of time, when $H=-\infty$, and an infinite avalanche only occurs if the value of $\mathbb{E}[\xi_i]$ increases to become greater than one. 
Whether this happens or not is defined by the sign of the derivative of $\mathbb{E}[\xi_i]$ with respect to time. 
The expression for $\dot{\mathbb{E}}[\xi_i]$, 
\begin{equation}
\dot{\mathbb{E}}[\xi_i]= A(t)\text{Cov}[\zeta_j,n_j]+B(t)~, 
\label{eq:E_xi}
\end{equation}
consists of two contributions associated 
with flip- ($\propto A(t)$) and field-induced ($\propto B(t)$)  changes (the functions $ A(t)$ and $B(t)$ are defined in App.~\ref{sec:DerivExi}).  
Here, 
\begin{equation}
\text{Cov}[\zeta_j,n_j]\equiv \mathbb{E}^{\text{F}}[n_j(\zeta_j-\mathbb{E}^{\text{F}}[\zeta_j])]~,\label{eq:Cov}
\end{equation}
and the function $B(t)>0$ remains finite at all times. 
In contrast, $A(t) >0$ diverges as $A(t)\propto (1-\mathbb{E}[\xi_i])^{-1}$, implying that the flip-induced mechanism is the main contribution to $\dot{\mathbb{E}}[\xi_i]$ when $\mathbb{E}[\xi_i]$ approaches 1 from below. 
In this case, $A(t) \gg 1$ and the system can only become supercritical if $\dot{\mathbb{E}}[\xi_i]>0$, meaning that the following condition must necessarily hold:
\begin{equation}
\lim_{\mathbb{E}[\xi_i] \to 1^-}\text{Cov}[\zeta_j,n_j]>0~.\label{eq:FinCond2}
\end{equation}
This criterion constitutes one of the main analytical findings of this paper. 

For a $q=3$ lattice, 
it follows from Eq.~\eqref{eq:Cov} that,
\begin{eqnarray}
&&\text{Cov}[\zeta_j,n_j]=\sum_{n_j=0}^2\sum_{\zeta_j=0}^2(1-\mathbb{E}^{\text{F}}[\zeta_j])n_jP^{\text{F}}(\zeta_j,n_j)\nonumber\\
&&=(1-\mathbb{E}^{\text{F}}[\zeta_j])P^{\text{F}}(1,1)-\mathbb{E}^{\text{F}}[\zeta_j]\left[P^{\text{F}}(0,1)+2P^{\text{F}}(0,2)\right]~,\nonumber\\\label{eq:Covq3}\end{eqnarray}
where $P^{\text{F}}(k,m)$ is the probability that $\zeta_j=k$ and $n_j=m$ (see App.~\ref{sec:DerivExi}).
In Eq.~\eqref{eq:Covq3}, we have used the property that $P^{\text{F}}(k,m)\neq 0$ only for $k=0$ or $k=q-1-m$ in order to cancel terms.
The only positive contribution ($\propto P^{\text{F}}(1,1)$), 
tends to zero as $\mathbb{E}^{\text{F}}[\zeta_j]\to 1^-$, while the negative contributions do not. As a result, $\text{Cov}[\zeta_j,n_j]$ becomes negative and condition~\eqref{eq:FinCond2} is not satisfied.

This illustration of the failure of the mechanism leading to an infinite avalanche for the $3$-regular graph provides the key to the qualitative understanding of the condition given by Eq.~\eqref{eq:FinCond2}.
Indeed, the main reason for the appearance of the infinite avalanche is that the interaction of an avalanche with pre-flipped spins (occurring at spins where $n_i\ge 1$) causes the avalanche to branch (i.e. $\zeta_i$ must be greater than one for this spin; see the evolution of the state of the central spin in Fig.~\ref{fig:avalanche}), making the BP super-critical. 
The infinite avalanche thus occurs when avalanche fronts are interacting with pre-flipped spins in a way which encourages branching.
This type of branching requires at least $3$ open paths (see paths $1$, $2$ and $3$ in Fig.~\ref{fig:avalanche}) from a flipping spin (one incoming (path $1$) and at least two outgoing (paths $2$ and $3$)) in addition to at least one path closed by the presence of the pre-flipped spins (path $4$). 
This means that an infinite avalanche driven by interaction with pre-flipped spins is only possible for $q\ge 4=[1$~(closed path)~$+1$~(incoming)~$+2$~(outgoing)$]$.

\begin{figure}
\includegraphics[height=10.0cm]{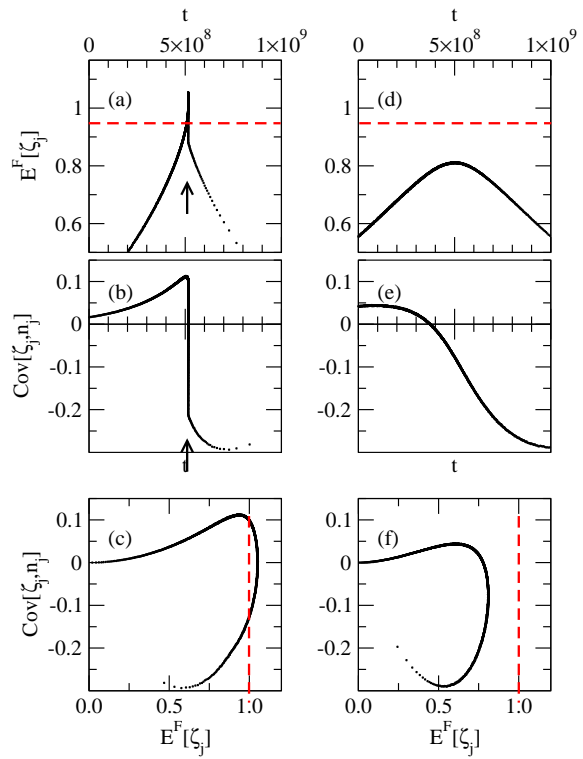}
\caption{The rate of branching $\mathbb{E}^{\text{F}}[\zeta_j]$ (a,d) and $\text{Cov}[\zeta_j,n_j]$ (b,e) {\it vs} time $t$. 
Panels (c) and (f) show $\text{Cov}[\zeta_j,n_j]$ {\it vs} $\mathbb{E}^{\text{F}}[\zeta_j]$ for data presented in (a)-(b) and (d)-(e), respectively. 
Data corresponds to 
a $4$-regular graph of $N=10^9$ spins, with normally distributed disorder below $\Delta=1.65$ (a,b,c), and above $\Delta=2.5$ (d,e,f), critical $\Delta_{\text{c}}\simeq 1.78215$. 
The external field was swept from $-\infty$ to $+\infty$ at a rate $\text{d}H/\text{d}t=2\times 10^{-9}$ with $H(t=0)=0$ and stochastic flips occurred at a rate $\Gamma=1$.
\label{fig:expAtCrit}
}
\end{figure}

Numerical support for condition~\eqref{eq:FinCond2} is presented in Fig.~\ref{fig:expAtCrit}.
The avalanches below (a,b,c) and above (d,e,f) criticality are analysed for a $4$-regular graph.
The infinite avalanche occurs stochastically after $\mathbb{E}^{\text{F}}[\zeta_j]$ (the rate of branching) passes $1$ (see (a)), at which time there is a non-zero probability of any given avalanche being infinite.
The positive derivative of $\mathbb{E}^{\text{F}}[\zeta_j]$ prior to the infinite avalanche is due to the fact that $\text{Cov}[\zeta_j,n_j]$ is positive when $\mathbb{E}^{\text{F}}[\zeta_j] \to 1^-$ (see (a) and (b)).
The infinite avalanche causes the covariance to reduce and eventually become negative, which, in turn leads to $\mathbb{E}[\xi_i]$ dropping below $1$ and the avalanche terminating (see (c)).
Above the critical degree of disorder ($\Delta>\Delta_{\text{c}}$), the branching rate grows initially and, despite approaching $1$ it is always $<1$ (see (d)). 
This is consistent with the fact that condition \eqref{eq:FinCond2} is not satisfied since $\text{Cov}[\zeta_j,n_j]$ becomes negative before $\mathbb{E}^{\text{F}}[\zeta_j]$ reaches its maximum value (cf. (d-f)).
These results suggest that monitoring the evolution of $\mathbb{E}^{\text{F}}[\zeta_j]$ and $\text{Cov}[\zeta_j,n_j]$ over relatively short time intervals prior to observing a large avalanche could be used to predict its occurrence.

\section{Time-dependent properties of avalanches} \label{sec:TimeDep}

Important characteristics of avalanches which can be measured experimentally~\cite{Kuntz2000,Mehta2002,SpasojevicEPL2006,Papanikolaou_NatPhys2011} include the distribution of avalanche durations, their power spectrum and pulse shapes.
Below, we demonstrate how, using the generating function techniques known for BPs~\cite{Jagers2005}, analytical expressions for these quantities can be derived.
These characteristics of avalanches can be defined in terms of the rate of change of magnetisation, $\text{d}\langle m\rangle/\text{d}t$ (where $\langle m\rangle=N^{-1}\left\langle \sum_is_i\right\rangle$), which is proportional to the experimentally measured voltage~\cite{Spasojevic1996} $\text{d}\langle m\rangle/\text{d}t\propto V(t)=V_0\sum_i\delta(t-t_i)$ (with $V_0$ being a coefficient of proportionality) where each $\delta$-function corresponds to the flip of a spin at time $t_i$.

In order to find these characteristics,
we consider an avalanche in which a set of spins, $\{s_i\}$, have flipped.
Within the continuous-time dynamics, these spins $s_i$ will flip at different times $t_i\ge 0$, with the first flip occurring at time $t=0$ and the final one at time $t=T$.
Let us assume that a single spin flips spontaneously at $t=0$ to initiate the avalanche (only one field-induced change).
When this spin flips, a random number, $\xi_i^{\text{I}}$, of its neighbours become unstable, which
is distributed according to $P^{\text{I}}(\xi_i^{\text{I}})$.
Assume that spin $i$ in the avalanche flips at time $t_i$.
This spin will also have a random number, $\xi_i$, of neighbours in the down-state which become unstable and this number is distributed according to $P^{\text{R}}(\xi_i)$.
Because $H(t)$ and thus $P^*$ vary adiabatically slowly with time, the values of $P^{\text{I}}(\xi_i^{\text{I}})$ and $P^{\text{R}}(\xi_i)$ remain unchanged during an avalanche. 
Accordingly, the spin-flip dynamics for any avalanche in the zt-RFIM can be mapped to a simpler form of BP, known as a continuous-time Galton-Watson (GW) process~\cite{Jagers2005}.
Therefore, we apply the generating function formalism developed for the GW process in order to describe the dynamical behaviour of the zt-RFIM. 
Note, however, that in general, both $P^{\text{I}}(\xi_i^{\text{I}})$ and $P^{\text{R}}(\xi_i)$ vary between different avalanches.

In order to use the generating function formalism, we introduce generating functions, $a(x)$ and $f(x)$, corresponding to the probability distributions $P^{\text{I}}(\xi_i^{\text{I}})$ and $P^{\text{R}}(\xi_i)$, i.e.
\begin{equation}
a(x)=\mathbb{E}[x^{\xi_i^{\text{I}}}]=\sum_{n=0}^qP^{\text{I}}(\xi_i^{\text{I}})x^{\xi_i^{\text{I}}}~,\label{eq:adef}
\end{equation}
and,
\begin{equation}
f(x)=\mathbb{E}[x^{\xi_i}]=\sum_{\xi_i=0}^qP^{\text{R}}(\xi_i)x^\xi_i~,\label{eq:fdef}
\end{equation}
where $P^{\text{I}}(\xi_i^{\text{I}})$ and $P^{\text{R}}(\xi_i)$ are given by Eqs.~\eqref{eq:ImmxiDist} and~\eqref{eq:RepxiDist}, respectively.
The number, $N^{\text{U}}$, of unstable spins in the avalanche at time $t$ is randomly distributed according to $P(N^{\text{U}},t)$, and
a generating function, $G_t(x)$, can also be written for this quantity,
\begin{equation}
G_t(x)=\mathbb{E}[x^{N^{\text{U}}}]=\sum_{N^{\text{U}}=0}^\infty P(N^{\text{U}},t)x^{N^{\text{U}}}~.\label{eq:Gtdef}
\end{equation}
Each spin $i$ unstable at a time $t$ will subsequently flip at time $t_i\ge t$.
This flip can cause its neighbours to become unstable, which will subsequently flip causing further spins to become unstable etc. 
In this way, there will be a total of $X_{i,t^\prime}$ unstable spins at a time $t+t^\prime$ emerging from a single initial unstable spin $i$ at time $t$.
The value of $X_{i,t^\prime}$ is 
randomly distributed with probability $P(X_{i,t^\prime})$, identical for all $i$ and independent of the time $t$.
The corresponding generating function is 
\begin{equation}
f_{t^\prime}(x)=\mathbb{E}[x^{X_{i,t^\prime}}]=\sum_{X_{i,t^\prime}=0}^\infty P(X_{i,t^\prime})x^{X_{i,t^\prime}}~,\label{eq:ftdef}
\end{equation}
and the value of $f_{t^\prime}(x)$ and its derivatives can be calculated using known techniques (see App.~\ref{sec:FFunction}).

Each of the $\xi_i^{\text{I}}$ spins $i$ becoming unstable after an immigration event at time $t=0$ leads to $X_{i,t}$ unstable spins at a subsequent time $t>0$ where each of the $X_{i,t}$ is distributed in the same way as $X_{i,t^\prime}$.
Thus the total number of unstable spins at a subsequent time $t>0$ is $N^\text{U}(t)=\sum_{i=1}^{\xi_i^{\text{I}}}X_{i,t}$ and its generating function can be written as,
\begin{eqnarray}
G_t(x)&=&\mathbb{E}[x^{N^\text{U}(t)}]=\mathbb{E}[x^{\sum_{i=1}^{\xi_i^{\text{I}}}X_{i,t}}]\nonumber\\
&=&\mathbb{E}[\left(\mathbb{E}[x^{X_{i,t}}]\right)^{\xi_i^{\text{I}}}]=a(f_t(x))~.\label{eq:GfromF}
\end{eqnarray}
Evaluation of the above generating functions and their derivatives is sufficient to calculate the values of $\rho_T$, $\langle V(t)\rangle_T$ and $S(\omega)$, as we do next.

\begin{figure} 
\includegraphics[width=8.5cm]{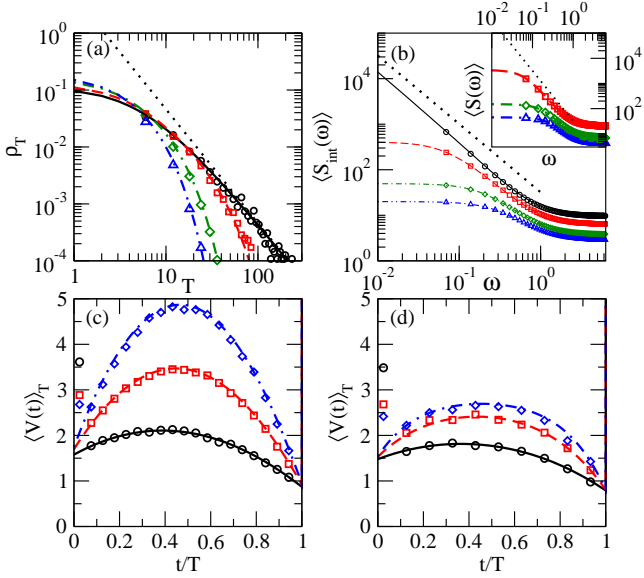}
\caption{Time-dependent properties of avalanches. 
Solutions of the exact equations are plotted with lines and compared with results from numerical simulations shown by symbols. 
Panels (a) and (b) show the distribution of avalanche durations, $\rho_T$, and the integrated power spectra, $\langle S_{\text{int}}(\omega)\rangle$, respectively, for a $4$-regular graph with normally distributed disorder, $\Delta=\Delta_{\text{c}}$ (critical disorder; solid lines and circles), $\Delta= 2.0$ (dashed lines and squares), $\Delta= 2.5$ (dot-dashed lines and diamonds) and $\Delta=3.0$ (double dot-dashed lines and triangles). 
The inset of panel (b) shows the power spectra, $\langle S(\omega)\rangle$ evaluated for the same parameters as for (a) and (b).
In both (a) and the inset of (b), $H=1.0$.
The solid line in (a) tends to the limit $\rho_T\propto T^{-2}$ for $T\gg \Gamma^{-1}$, shown by the dotted line.
In (b) and its inset, the dotted lines indicate $S_{\text{int}}(\omega)\propto\omega^{-3/2}$ and $S(\omega)\propto \omega^{-2}$, respectively, corresponding to the asymptotic behaviour for $\omega\ll\Gamma$ near critical disorder.
Panels (c) and (d) show the shape of avalanches $\langle V(t)\rangle_T$ vs $t/T$ for $\Delta=\Delta_{\text{c}}$ and $\Delta=2.5$, respectively, $H=1.0$ and $T=10$ (solid lines and circles), $T=20$ (dashed lines and squares) and $T=30$ (dot-dashed lines and diamonds). 
Numerical data for durations and spectra were averaged over avalanches occurring when, $0.998<H<1.002$ ((a) and inset of (b)) and $0.98<H<1.02$ ((c) and (d)) for $10^3$ realisations of $10^6$ spins and $\Gamma=V_0=1$.
\label{fig:Dynamics}}
\end{figure}

\subsection{Avalanche Durations}

The distribution of the duration of avalanches, $\rho_T$, can be found using $G_T(0)=P(N^{\text{U}}=0)$, where $P(N^{\text{U}}=0)$ is the probability that an avalanche is extinct at time $T$. 
The distribution of avalanche durations is the derivative of this quantity,
\begin{equation}
\rho_T=\frac{\text{d}G_T(0)}{\text{d}T}=\partial_xa(f_T(0))\partial_Tf_T(0)~.\label{eq:Duration}
\end{equation}
Here, $\partial_xa(x_0)$ refers to the first derivative of $a(x)$ with respect to $x$ evaluated at $x=x_0$, while $\partial_Tf_T(x)=(\partial/\partial T) f_T(x)$.
Numerical evaluation of $\partial_Tf_T(0)$ following the methods described in App.~\ref{sec:FFunction} leads to the results for $\rho_T$ shown with lines in Fig.~\ref{fig:Dynamics}(a). As can be seen, the obtained curves are in excellent agreement with numerical simulations (symbols in Fig.~\ref{fig:Dynamics}(a)). The exact solution of $\rho_T$ is supported by analytical approximations obtained for large $T$. 
When the zt-RFIM is away from its critical point and is not close to snapping, the BP describing the avalanches is in its sub-critical regime.
Thus, in the limit of $T\gg \Gamma^{-1}$, $\rho_T$ decays exponentially according to (see App.~\ref{sec:FFunction}),
\begin{equation}
\rho_T\propto \exp[-\Gamma T \mathbb{E}[1-\xi_i]]~.\label{eq:rhotsubcrit}
\end{equation}
Conversely, at criticality in the zt-RFIM, the BP is also in its critical regime, meaning that at large times, $\rho_T$ obeys a power law (see App.~\ref{sec:FFunction}), i.e.
\begin{equation}
\rho_T\simeq \frac{2 \mathbb{E}[\xi_i^{\text{I}}]}{\mathbb{E}[\xi_i(\xi_i-1)]}\Gamma^{-1} T^{-2}~,\label{eq:rhotcrit}
\end{equation}
for $T\gg \Gamma^{-1}$. 
Such a power law dependence on $T$ (see dotted line in Fig.~\ref{fig:Dynamics}(a)) is the same as the one found by the mean-field approach~\cite{Dahmen1995,Durin2000,Durin2001,SpasojevicEPL2006}.

\subsection{Avalanche Spectrum}

The power spectrum of avalanches, defined as the mean square of the Fourier transform of the voltage
can be written as,
\begin{eqnarray}
S(\omega)&=&\left\langle\left|\int_{t=-\infty}^\infty e^{i\omega t}V(t)\text{d}t\right|^2\right\rangle\nonumber\\
&=&\int_{t=-\infty}^\infty\int_{t^\prime=-\infty}^\infty e^{i\omega (t-t^\prime)}\left\langle V(t)V(t^\prime)\right\rangle\text{d}t\text{d}t^\prime~,\label{eq:SpectrumC}
\end{eqnarray}
where the angular brackets refer to averaging over all avalanches occurring for external field in the range $[H,H+\delta H]$, with $\delta H\to 0$.
The first spin flips deterministically at $t=0$ and must be treated separately from the other stochastically flipping spins.
Therefore, we define $U(t)=V_0\sum_{i\neq 1}\delta(t-t_i)$ as the voltage created by spin flips except the first one, then $\left\langle V(t)V({t^\prime})\right\rangle$ contains four terms as follows,
\begin{eqnarray}
\left\langle V(t)V({t^\prime})\right\rangle&=&V_0^2\delta(t)\delta(t^\prime)+V_0\delta(t^\prime)\left\langle U(t)\right\rangle\nonumber\\&+&V_0\delta(t)\left\langle U(t^\prime)\right\rangle+\left\langle U(t)U({t^\prime})\right\rangle~.\label{eq:TotVoltAutoCorr}
\end{eqnarray}

The mean voltage $\left\langle U(t)\right\rangle$ is equal to $V_0$ multiplied by the probability that a spin flips in a time interval $[t,t+\delta t]$,
\begin{equation}
\left\langle U(t)\right\rangle=V_0\text{Prob}(\text{Flip in }[t,t+\delta t])(\delta t)^{-1}~.\label{eq:meanU}
\end{equation}
The mean $\left\langle U(t)U(t^\prime)\right\rangle$ contains two terms corresponding to the cases $t=t^\prime$ and $t\neq t^\prime$, i.e.
\begin{eqnarray}
&&\left\langle U(t)U(t^\prime)\right\rangle\nonumber\\&=& 
V_0^2\text{Prob}(\text{Flip in }[t,t+\delta t])\delta(t-t^\prime)(\delta t)^{-1}\nonumber \\ &+&
V_0^2\text{Prob}(\text{Flip in }[t,t+\delta t]\cap\text{ Flip in }[t^\prime,t^\prime+\delta t])(\delta t)^{-2}~.\nonumber\\\label{eq:autoCorr}
\end{eqnarray}
In Eqs.~\eqref{eq:meanU} and~\eqref{eq:autoCorr}, $\text{Prob}(\text{Flip in }[t,t+\delta t])$ is given by,
\begin{equation}
\text{Prob}(\text{Flip in }[t,t+\delta t])=\mathbb{E}[\Gamma\delta t N^{\text{U}}(t)]=\Gamma\partial_xG_t(1)\delta t~,
\end{equation}
and the probability that a spin flips in the interval $[t,t+\delta t]$ and another spin flips in the interval $[t^\prime,t^\prime+\delta t]$, can be found as,
\begin{eqnarray}
&&\text{Prob}(\text{Flip in }[t,t+\delta t]\cap\text{ Flip in }[t^\prime,t^\prime+\delta t])\nonumber\\
&=&\left\langle \Gamma\delta t N^{\text{U}}(t_{\text{min}})\sum_{i=1}^{N^{\text{U}}(t_{\text{min}})+\xi_i-1}\Gamma\delta t X_{|t-t^\prime|,i}\right\rangle\nonumber\\
&=&\Gamma^2(\delta t)^2\left(\mathbb{E}[ N^{\text{U}}(t_{\text{min}})(N^{\text{U}}(t_{\text{min}})-1)]\right.\nonumber\\&&\left.+\mathbb{E}[ N^{\text{U}}(t_{\text{min}}) ] \mathbb{E}[\xi_i ]\right) \mathbb{E}[ X_{|t-t^\prime|} ]\nonumber\\
&=&
 \Gamma^2(\delta t)^2[\partial_{xx}G_{t_{\text{min}}}(1)+\partial_xG_{t_{\text{min}}}(1)\partial_xf(1)]\partial_xf_{|t-t^\prime|}(1)~,\nonumber\\
\label{eq:VoltCrossCorr} 
\end{eqnarray}
where $t_{\text{min}}=\min(t,t^\prime)$.
The factor $\Gamma\delta t N^{\text{U}}(t_{\text{min}})$, in the second line of Eq.~\eqref{eq:VoltCrossCorr}, refers to the probability that a spin flips at the earlier of the two times $t$ and $t^\prime$. 
After this spin flips, with $\xi_i$ offspring, there are $N^{\text{U}}(t_{\text{min}})+\xi_i-1$ unstable spins.
Each of these unstable spins gives rise to $X_{|t-t^\prime|,i}$ unstable spins at the later of the two times $t$ and $t^\prime$, meaning that probability of a spin flip at the later time is,
$\sum_{i=1}^{N^{\text{U}}(t_{\text{min}})+\xi_i-1}\Gamma\delta t X_{|t-t^\prime|,i}$.

Substituting Eqs.~\eqref{eq:TotVoltAutoCorr}-\eqref{eq:VoltCrossCorr} back into Eq.~\eqref{eq:SpectrumC} results in,
\begin{widetext}
\begin{eqnarray}
S(\omega)&=&V_0^2\left(1+2\Gamma\partial_xa(1)\int_{t=0}^{\infty}\exp\left[-\Gamma t\mathbb{E}[ 1-\xi_i ]\right]\cos\left(\omega t\right)\text{d}t+\Gamma\partial_xa(1)\int_{t=0}^{\infty}\exp\left[-\Gamma t \mathbb{E}[ 1-\xi_i ]\right]\text{d}t\right.\nonumber\\
&&\left.+\Gamma^2\int_{t=0}^{\infty}\left(\partial_{xx}G_{t}(1)+\partial_xG_{t}(1)\partial_xf(1)\right)\int_{u=0}^{\infty}\exp\left[-\Gamma u \mathbb{E}[ 1-\xi_i ]\right]\cos\left(\omega u\right)\text{d}u\text{d}t\right)\nonumber\\
&=&S_0+\frac{S_1}{\mathbb{E}[\xi_i-1 ]^2+\omega^2\Gamma^{-2}}~,\label{eq:SpectrumFinal}
\end{eqnarray}
\end{widetext}
where,
\begin{eqnarray}
S_0&=&V_0^2\left[1+\frac{\mathbb{E}[\xi_i^{\text{I}}]}{\mathbb{E}[ 1-\xi_i]}\right]~,\label{eq:S0}
\end{eqnarray}
and,
\begin{eqnarray}
S_1&=&V_0^2\left[\mathbb{E}[\xi_i^{\text{I}}]\left(\frac{\mathbb{E}[\xi_i(\xi_i-1)]}{\mathbb{E}[ 1-\xi_i ]}+1\right)+\mathbb{E}[\left(\xi_i^{\text{I}}\right)^2]\right]~.\label{eq:S1}
\end{eqnarray}
We have simplified Eqs.~\eqref{eq:S0} and~\eqref{eq:S1} using the relationships $\partial_xa(1)=\mathbb{E}[\xi_i^{\text{I}}]$ and $\partial_{xx}a(1)+\partial_xa(1)= \mathbb{E}[\left(\xi_i^{\text{I}}\right)^2]$ where these moments of $\xi_i^{\text{I}}$ are given by,
\begin{eqnarray}
\mathbb{E}[\xi_i^{\text{I}}]&=&\left(\frac{\partial F_0}{\partial P^*}\right)\left(\frac{\partial F_1}{\partial H}\right)\left(\frac{\partial F_0}{\partial H}\right)^{-1}\nonumber\\
\mathbb{E}[\left(\xi_i^{\text{I}}\right)^2 ]&=&\left(1-q^{-1}\right)\left(\frac{\partial F_0}{\partial P^*}\right)^2\left(\frac{\partial F_2}{\partial H}\right)\left(\frac{\partial F_0}{\partial H}\right)^{-1}+\mathbb{E}[\xi_i^{\text{I}}]~.\nonumber\\
\end{eqnarray}
Except for the constant factor $S_0$, the functional form of the power spectrum given by Eq.~\eqref{eq:SpectrumFinal} is the same as that found within the mean-field approach~\cite{Kuntz2000,SpasojevicEPL2006}. 
Near the critical point, the term $\mathbb{E}[1-\xi_i]$ goes to zero according to a power law, $\mathbb{E}[1-\xi_i]\propto \left|H-H_{\text{C}}\right|^{2/3}$\cite{Handford2012}, while all other expectation values involving $\xi_i$ and $\xi_i^{\text{I}}$ in Eqs.~\eqref{eq:SpectrumFinal}-\eqref{eq:S1} are constant to leading order.
This means that near criticality and for small values of $\omega$, the spectrum is given by,
\begin{equation}
S(\omega)\sim \left|H-H_{\text{C}}\right|^{-2/3}\left(\frac{1}{A\left|H-H_{\text{C}}\right|^{4/3}+\omega^2\Gamma^{-2}}\right)~,\label{eq:CritSpec}
\end{equation}
so that at criticality, the spectrum can be written $S(\omega)\propto S_1(H)\omega^{-2}$ (see Fig.~\ref{fig:Dynamics}(b))~\cite{Cote1991,Kuntz2000}.
The divergence in $S(\omega)$ at $\omega=0$ can be integrated out considering the average of $S(\omega)$ for all avalanches occurring in a single magnetisation reversal, i.e.
\begin{equation}
S_{\text{int}}(\omega)=\int_{t=-\infty}^{\infty}S(\omega,H(t))R^{\text{I}}\text{d}t~.\label{eq:IntSpec}
\end{equation}
This quantity also follows a power law at criticality, $\langle S_{\text{int}}(\omega)\rangle\propto\omega^{-\theta}$, with $\theta=3/2$ (see inset of Fig.~\ref{fig:Dynamics}(b)). 
This result follows by substituting Eq.~\eqref{eq:CritSpec} into the integrand in Eq.~\eqref{eq:IntSpec}.
In fact, 
experimental measurements of the spectrum~\cite{Spasojevic1996,Mehta2002} involve a sum across avalanches near the critical point, and this exponent can be compared with the experimentally observed exponent in the range $\theta\simeq 1.6-1.7$~\cite{Dahmen1995,Spasojevic1996}.

\subsection{Pulse Shapes}

The pulse shapes of avalanches can be defined as,
\begin{equation}
\langle V(t)\rangle_T=\left\langle V_0\sum_{i}\delta(t-t_i)\right\rangle_T~,\text{ for }0\le t\le T
\end{equation} 
where the mean is taken over all avalanches with a duration in a small range of values
$[T,T+\delta T]$. 
This definition is consistent with those used in the discrete time formulation~\cite{Mehta2002} and experimental setups~\cite{Spasojevic1996}.
In order to average the pulse shape over all avalanches of fixed duration $T$, 
we need to calculate the probability that a spin flips within a time interval $[t,t+\delta t]$, given that an avalanche goes extinct at time $T$.
Let $N^{\text{U}}(t)$ be the number of unstable spins at time $t$.
The probability that one of these spins flips in an interval $[t,t+\delta t]$ (where $\delta t\to 0$) is given by,
\begin{equation}
\text{Prob}(\text{Flip in }[t,t+\delta t]|N^{\text{U}}(t))=\Gamma N^{\text{U}}(t)\delta t~. \label{eq:PFlipTime}
\end{equation}
If a spin does flip it will destabilise $\xi_i$ new spins and then there will be $N^{\text{U}}(t)+\xi_i-1$ unstable spins in the system.
After this event, each unstable spin causes a sub-avalanche of spin flips which goes extinct before time $T$ with probability $f_{T-t}(0)$.
The probability that all of these sub-avalanches will be extinct before time $T$ is therefore,
\begin{eqnarray}
&&\text{Prob}(\text{Ext} < T|\text{Flip in }[t,t+\delta t]\cap N^{\text{U}}(t)\cap\xi_i)\nonumber\\&&=\left[f_{T-t}(0)\right]^{N^{\text{U}}(t)+\xi_i-1}~.
\end{eqnarray}
The probability that the last of these avalanches goes extinct in an interval $[T,T+\delta T]$ (where $\delta T\to 0$) is,
\begin{eqnarray}
&&\text{Prob}(\text{Ext in }[T,T+\delta T]|\text{Flip in }[t,t+\delta t]\cap N^{\text{U}}(t)\cap\xi_i)\nonumber\\&&=\delta T\frac{\partial}{\partial T} \left[f_{T-t}(0)\right]^{N^{\text{U}}(t)+\xi_i-1}~.\label{eq:PExtinctGivenFlip}
\end{eqnarray}
Combining Eqs.~\eqref{eq:PFlipTime} and~\eqref{eq:PExtinctGivenFlip}, the probability that a spin flips in the interval $[t,t+\delta t]$ given that the avalanche goes extinct in an interval $[T, T+\delta T]$ can be calculated as,
\begin{eqnarray}
&&\text{Prob}(\text{Flip in }[t,t+\delta t]|\text{Ext in }[T,T+\delta T])\nonumber\\&&={\rho_T}^{-1}\left\langle \Gamma N^{\text{U}}(t)\frac{\partial} {\partial T}\left[f_{T-t}(0)\right]^{N^{\text{U}}(t)+\xi_i-1}\right\rangle\delta t~,\nonumber\\
\end{eqnarray}
where the angular brackets indicate an average over the distributions $P(N^{\text{U}},t)$ and $P(\xi_i)$ of $N^{\text{U}}(t)$ and $\xi_i$, respectively (see Eqs.~\eqref{eq:fdef} and~\eqref{eq:Gtdef}). 
The mean pulse shape is $V_0$ multiplied by the rate at which spins flip, i.e. $\langle V(t)\rangle_T$ is given by, 
\begin{eqnarray}
&&\langle V(t)\rangle_T\nonumber\nonumber\\&=&V_0\text{Prob}(\text{Flip in }[t,t+\delta t]|\text{Ext in }[T,T+\delta T])(\delta t)^{-1}\nonumber\\
&=&V_0{\rho_T}^{-1}\left\langle \Gamma N^{\text{U}}(t)\frac{\partial} {\partial T}\left[f_{T-t}(0)\right]^{N^{\text{U}}(t)+\xi_i-1}\right\rangle\nonumber\\&&+V_0\delta(t)+V_0\delta(T-t)~,\label{eq:PulseC}
\end{eqnarray}
where the delta-functions $\delta(t)$ and $\delta(T-t)$ account for the first and last spin flips which occur at $t=0$ and $t=T$, making the above equation valid for $0\le t\le T$.
In order to perform the average over the distributions of $N^{\text{U}}$ and $\xi_i$, Eq.~\eqref{eq:PulseC} can be written in terms of the generating functions $f(x)$ and $a(x)$ defined above,
\begin{widetext}
\begin{eqnarray}
\langle V(t)\rangle_T\nonumber&=&V_0\Gamma{\rho_T}^{-1}\frac{\partial^2 }{\partial T\partial x}\left[ \mathbb{E}[(xf_{T-t}(0))^{N^{\text{U}}(t)}]\mathbb{E}[ f_{T-t}(0)^{\xi_i} ]\left(f_{T-t}(0)\right)^{-1}\right]_{x=1}+V_0\delta(t)+V_0\delta(T-t)\nonumber\\
&=&V_0\Gamma{\rho_T}^{-1}\frac{\partial^2 }{\partial T\partial x}\left[G_t(xf_{T-t}(0))f(f_{T-t}(0))\left(f_{T-t}(0)\right)^{-1}\right]_{x=1}+V_0\delta(t)+V_0\delta(T-t)\nonumber\\
&=&V_0\Gamma\left(\frac{\partial_{xx}a(f_T(0))}{\partial_{x}a(f_T(0))}\partial_{x}f_t(f_{T-t}(0))+\frac{\partial_{xx}f_t(f_{T-t}(0))}{\partial_xf_t(f_{T-t}(0))}\right)f(f_{T-t}(0))\nonumber\\
&&+V_0\Gamma\partial_{x}f(f_{T-t}(0))+V_0\delta(t)+V_0\delta(T-t)~,\label{eq:PulseShape}
\end{eqnarray}
\end{widetext}
where we have simplified the second line in the above expression using Eqs.~\eqref{eq:GfromF} and~\eqref{eq:Duration} and the identities $f_t(f_{T-t}(0))=f_T(0)$ and $\partial_Tf_T(0)=\partial_xf_t(f_{T-t}(0))\partial_Tf_{T-t}(0)$. 
In order to evaluate $\langle V(t)\rangle_T$ using Eq.~\eqref{eq:PulseShape}, the coefficients of the polynomials $f(x)$ and $a(x)$ should be found employing Eqs.~\eqref{eq:adef} and \eqref{eq:fdef}.
Then the values of $f_{T-t}(0)$, $f(f_{T-t}(0))$ and their derivatives along with the derivatives of $a(f_T(0))$ and $f_t(f_{T-t}(0))$ can be found 
as outlined in App.~\ref{sec:FFunction}. 

\begin{figure}
\includegraphics[height=5.5cm]{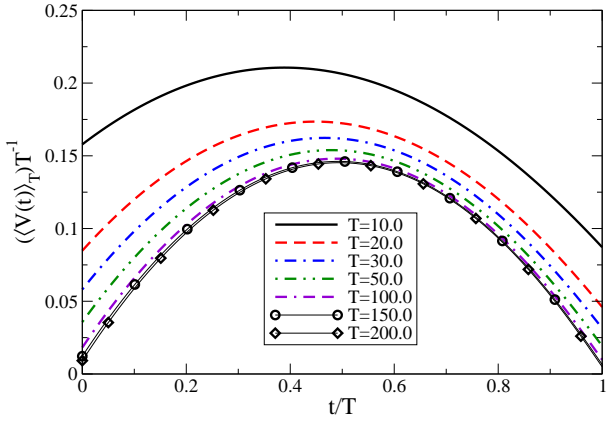}
\caption{Scaling function ${\hat{V}}(t/T)$ in the scaling hypothesis for the pulse shape, $\langle V(t)\rangle_T=T^{x} {\hat{V}}(t/T)$ with $x=1$, suggested in Ref.~\cite{Sethna2001,Baldassarri2003}.  
The pulse shape was calculated for a $4$-regular graph with normal disorder at criticality ($\Delta=\Delta_{\text{c}}$, $H=1.0$) using Eq.~\eqref{eq:PulseShape}. 
Different curves represent different avalanche durations as marked in the figure legend, with $\Gamma=V_0=1$.
\label{fig:ScalingPulse}}
\end{figure}

At criticality, the pulse shape given by Eq.~\eqref{eq:PulseShape} is similar to that for a simple random walk (see Fig.~\ref{fig:Dynamics}(c)), while away from criticality, the pulse shape flattens for large $T$, similar to the effect found for a 1D random walk in a parabolic potential well~\cite{Baldassarri2003} (see Fig.~\ref{fig:Dynamics}(d)).
At $\Delta=\Delta_\text{c}$, the pulse shape is significantly asymmetric for $T \lesssim 100$ and a scaling hypothesis~\cite{Kuntz2000,Sethna2001,Mehta2002,Papanikolaou_NatPhys2011} of the form $\langle V(t) \rangle_T = T^x \hat{V}(t/T)$ is only appropriate for durations $T \gtrsim 100$, as can be seen from Fig.~\ref{fig:ScalingPulse}. 
It can be noted that the distribution $\rho_T$ (see Fig.~\ref{fig:Dynamics}(a)) also deviates significantly from the power-law behaviour for short avalanches. 
Therefore, experimental studies might not reliably detect power-law scaling of dynamical properties unless they focus on very long avalanches, with the drawback that they are much less frequent.

\section{Conclusions} \label{sec:Conclusions}

The mechanisms of evolution of avalanches in the zero-temperature random field Ising model on a $q$-regular graph have been found in terms of a mapping to a branching process describing an evolving population with immigration and reproduction followed by immediate death.
The global avalanche response has been linked to local (microscopic) quantities such as the mean number $\mathbb{E}[\xi_i]$ of spins becoming unstable after the flip of a neighbour and the interaction between the propagating front of an avalanche and spins flipped in previous avalanches (formally accounted by $\text{Cov}[\zeta_j,n_j]$).
From this, we have answered the question as to why the jump in magnetisation does not occur in Bethe lattices with $q \le 3$~\cite{Sabhapandit2000,SabhapanditPRL}, but does occur for $q>3$.
The mapping we have established sheds light on the underlying mechanisms which cause an infinite avalanche on the $q$-regular graph to occur.

The mapping of our model to the branching process allowed the generating function formalism to be applied for analytical derivation of several important experimentally measurable quantities, including the distributions of avalanche durations, avalanche pulse-shapes, and power spectra.
The scaling laws obtained for these quantities measured on $q$-regular graphs fit well with the behaviour known from the mean-field approach~\cite{Sethna2001,Kuntz2000}.
The scaling for avalanche duration and pulse-shapes is found to be valid only for rare avalanches of large duration.

Our results are based on a relatively simple framework but they serve as a solid mathematical foundation for future work aiming at elucidating the evolutionary mechanisms of avalanches in more realistic systems.
In particular, our analysis suggests that monitoring the evolution of appropriate local quantities can be used to predict the occurrence of infinite avalanches. 
The prediction techniques do not rely on the monitoring of system-wide quantities (e.g. susceptibility~\cite{Acharyya1996,Pradhan2001}) or time-dependent quantities (e.g. a minimum in failure rate~\cite{Pradhan2009,Pradhan2011}), which may be more difficult to measure for various types of phenomena.

\section{Acknowledgements}

TPH acknowledges the financial support of the EPSRC.

\appendix 

\section{Derivation of $\text{d}\mathbb{E}[\xi_i]/\text{d}t$}\label{sec:DerivExi}

In this section, we derive the expression for $\text{d}\mathbb{E}[\xi_i]/\text{d}t$ used in Sec.~\ref{sec:Condition}.
It follows from Eqs.~\eqref{eq:SelfConsistent} and~\eqref{eq:RepxiDist} that,
\begin{eqnarray}
&&\mathbb{E}[\xi_i]=\frac{\partial F_1(P^*,H(t))}{\partial P^*}\nonumber\\&&=\frac{1-P^*}{1-F_1(P^*,H(t))}\left(\frac{\partial F_1(P^*,H(t))}{\partial P^*}\right)=\nonumber\\
&&\frac{\sum_{n=0}^{q-1}{\binom{q-1}{n}}(q-1-n)\left(P^*\right)^n\left(1-P^*\right)^{q-1-n}(p_{n+1}-p_n)}{\sum_{n=0}^{q-1}{\binom{q-1}{n}}\left(P^*\right)^n\left(1-P^*\right)^{q-1-n}(1-p_n)}~.\nonumber\\\label{eq:Meanxi}
\end{eqnarray}
The derivative of $\mathbb{E}[\xi_i]$ can be evaluated as,
\begin{eqnarray}
\frac{\text{d}\mathbb{E}[\xi_i]}{\text{d}t}&=&\left(\frac{\partial \mathbb{E}[\xi_i]}{{\partial P^*}}\right)\left(\frac{\text{d}P^*}{\text{d}t}\right)+\left(\frac{\partial \mathbb{E}[\xi_i]}{\partial H}\right)\left(\frac{\text{d}H}{\text{d}t}\right)~,\label{eq:dExi}
\end{eqnarray}
where the derivative of $P^*$ can be found using Eq.~\eqref{eq:SelfConsistent} as,
\begin{eqnarray}
\frac{\text{d}P^*}{\text{d}t}&=&\left(1-\frac{\partial F_1(P^*,H(t))}{\partial P^*}\right)^{-1}\nonumber\\&\times&\left(\frac{\partial F_1(P^*,H(t))}{\partial H}\right)\left(\frac{\text{d}H}{\text{d}t}\right)~,\label{eq:dPstar}
\end{eqnarray}
which is discontinuous only at $\partial F_1(P^*,H(t))/\partial P^*\to 1^-$.
Combining Eqs.~\eqref{eq:Meanxi},~\eqref{eq:dExi} and~\eqref{eq:dPstar} we obtain,
\begin{widetext}
\begin{equation}
\frac{\text{d}\mathbb{E}[\xi_i]}{\text{d}t}=A(t)\left[\left(\sum_{\zeta_i=0}^{q-1}\sum_{n_i=0}^{q-1}P^{\text{F}}(n_j,\zeta_j)\zeta_in_i\right)-\left(\sum_{\zeta_i=0}^{q-1}\sum_{n_i=0}^{q-1}P^{\text{F}}(n_i,\zeta_i)\zeta_i\right)\left(\sum_{\zeta_i=0}^{q-1}\sum_{n_i=0}^{q-1}P^{\text{F}}(n_j,\zeta_j)n_i\right)\right] +B(t)~,\label{eq:MeanxiDeriv}
\end{equation}
where the function,
\begin{equation}
B(t)=\left[\left(\frac{\partial^2F_1(P^*,H(t))}{\partial H\partial P^*}\right)+\left(1-P^*\right)^{-1}\left(\frac{\partial F_1(P^*,H(t))}{\partial H}\right)\left(\frac{\partial F_1(P^*,H(t))}{\partial P^*}\right)\right]\frac{\text{d}H}{\text{d}t}~,
\end{equation}
remains finite at all times. 
\end{widetext}
In contrast, the function,
\begin{eqnarray}
A(t)&=&\left(1-\mathbb{E}[\xi_i]\right)^{-1}\left(P^*\left(1-P^*\right)\right)^{-1}\nonumber\\&\times&\left(\frac{\partial F_1(P^*,H(t))}{\partial H}\right)\left(\frac{\text{d}H}{\text{d}t}\right)~,
\label{eq:Aoft}
\end{eqnarray}
diverges, $A(t)\to+\infty$, as $\mathbb{E}[\xi_i]\to 1^-$ because all four factors in Eq.~\eqref{eq:Aoft} are positive (see Eq.~\eqref{eq:Fm}).
The probability distribution $P^{\text{F}}(n_j,\zeta_j)$ in Eq.~\eqref{eq:MeanxiDeriv} is defined by,
\begin{equation}
P^{\text{F}}(n_j,\zeta_j)=P^{\text{R}}(\zeta_i)P^{\text{SU}}\delta_{\zeta_i,q-1-n_i}+P^{\overline{\text{R}}}(n_i)(1-P^{\text{SU}})\delta_{\zeta_i,0}~.\label{eq:MeanFCalc}
\end{equation}
Here, the quantity $P^{\overline{\text{R}}}(n_i)$, defined as, 
\begin{eqnarray}
P^{\overline{\text{R}}}(n_i)&=&{\binom{q-1}{n_i}}\left(P^*\right)^{n_i}\left(1-P^*\right)^{q-1-n_i}(1-p_{n_i+1})\nonumber\\&\times&\left[\left(1-P^*\right)\left(1-P^{\text{SU}}\right)\right]^{-1}~,
\end{eqnarray}
gives the probability that spin $i$, which remains stable when one of its neighbours changes from U to F, has $n_i$ flipped neighbours in addition to the one which changed state. 
Note that reproduction does not occur in this process at site $i$, indicated by $\overline{\text{R}}$).
The product of probabilities $P^{\text{R}}(\zeta_i)P^{\text{SU}}$ in Eq.~\eqref{eq:MeanFCalc} is the probability that a spin $i$ becomes unstable and has $\zeta_i$ stable neighbours when one of its neighbours moves from state U$\to$F.
The Kronecker-$\delta$'s in Eq.~\eqref{eq:MeanFCalc} ensure that $\zeta_i=q-1-n$ when spin $i$ becomes unstable and $0$ when spin $i$ remains stable.
The probability distribution $P^{\text{F}}(n_j,\zeta_j)$ is therefore the joint distribution of $n_i$ and $\zeta_i$ for any stable spin for which a neighbour changes state from U$\to$F, meaning that the term in the square brackets in Eq.~\eqref{eq:MeanxiDeriv} is an expression for the covariance,
\begin{equation}
\text{Cov}[\zeta_j,n_j]=\mathbb{E}^{\text{F}}[\zeta_jn_j]-\mathbb{E}^{\text{F}}[\zeta_j]\mathbb{E}^{\text{F}}[n_j]~.
\end{equation}
used in Sec.~\ref{sec:Condition}.
Comparison of Eqs.~\eqref{eq:RepxiDist} and~\eqref{eq:MeanFCalc} reveals the formula $\mathbb{E}[\xi_i]=\mathbb{E}^{\text{F}}[\zeta_j]$, used in Sec.~\ref{sec:Dynamics}.

\section{Calculation of $f_t(x)$}\label{sec:FFunction}

\begin{figure}
\includegraphics[height=6.0cm]{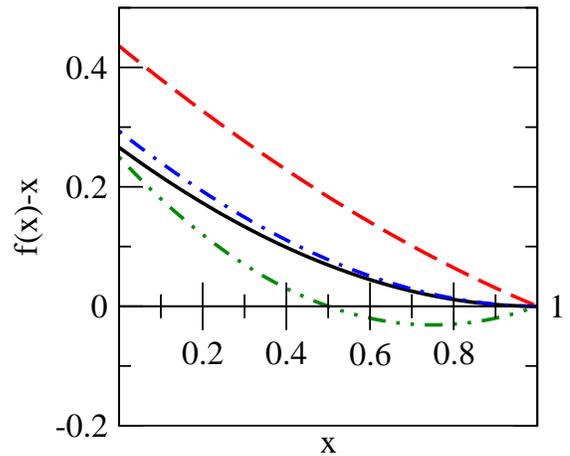}
\caption{Denominator of integral in Eq.~\eqref{eq:GWSolution} {\it vs} $x$ for high disorder, $\Delta=3.0$ and $H=1.0$ (red-dashed line, sub-critical regime of BP), at the critical point of the zt-RFIM characterised by $\Delta_{\text{c}}=1.78215$ and $H_{\text{c}}=1.0$ (black solid line, critical regime of BP), and immediately before snapping at low disorder $\Delta=1.0$ and the coercive field $H_{\text{coer}}(\Delta=1.0)=1.44754$ (blue dot-dashed line, critical regime of BP). 
The green double-dot dashed line would correspond to a hypothetical super-critical regime of the BP process but cannot be achieved for any values of $H$ or $\Delta$. 
\label{fig:fFunction}}
\end{figure}

In this appendix, we calculate the properties of the generating function $f_t(x)$ for the distribution of the number of unstable spins in a GW process as a function of time.
Using techniques previously applied to the GW process~\cite{Jagers2005} the generating function $f_t(x)$ can be found to obey,
\begin{equation}
\frac{\text{d}f_t(x)}{\text{d}t}=\Gamma[f(f_t(x))-f_t(x)],~~~f_0(x)=x~,\label{eq:G_ODE}
\end{equation}
where $f(x)$ is the generating function for the number of offspring of the individuals produced by reproduction.
Eq.~\eqref{eq:G_ODE} implies that $f_t(x)$ is the solution of the following equation, 
\begin{equation}
\int_{x^\prime =x}^{f_t(x)} \frac{1}{f(x^\prime)-x^\prime}\text{d}x^\prime=\Gamma t~.\label{eq:GWSolution}
\end{equation}
if $f(x)-x\neq 0$, and $f_t(x)=x$ if 
\begin{equation}
f(x)-x=0~\label{eq:rootsfx}
\end{equation}
The function $f(x)$ takes the value $1$ for $x=1$ and is convex in the interval $[0,1]$ since it is a polynomial with positive coefficients, $P^{\text{R}}(\xi_i)$ (see Eq.~\eqref{eq:fdef}). 
Therefore, Eq.~\eqref{eq:rootsfx} can have either one non-degenerate solution, $x=1$ (stable, $\partial_xf(1)<1$), a single degenerate solution at $x=1$ (stable for perturbations described by $x<1$, $\partial_xf(1)=1$, $\partial_{xx}f(1)>0$) or two solutions, $x=Q<1$ (stable, $\partial_xf(Q)<1$) and $x=1$ (unstable, $\partial_xf(1)>1$). 
This implies that there could be three possible regimes for the non-stationary solutions of $f_t(x)$ given by Eq.~\eqref{eq:GWSolution}, sub-critical, critical and super-critical, respectively, which correspond to popping, crackling and snapping behaviour in the zt-RFIM (see Fig.~S\ref{fig:fFunction}). 
It should be mentioned that the critical regime for the BP occurs not only at the critical point, $\Delta=\Delta_{\text{c}}$ and $H=H_{\text{c}}$, of the zt-RFIM.
In fact, the BP enters the critical regime as the external field approaches the coercive field at which snapping occurs, $H_{\text{coer}}(\Delta)$, for any disorder $\Delta<\Delta_{\text{c}}$.
The function $f(x)-x$ is given by Eq.~\eqref{eq:fdef} and it is non-negative for $0\le x\le 1$ for all values of $H$ or $\Delta$ (see Fig.~S\ref{fig:fFunction}). 
This means that the BP cannot be in the super-critical regime.
Indeed, any small increase in $H$ from $H=H_{\text{coer}}(\Delta)$ will cause the BP to enter the super-critical regime but an infinite avalanche (corresponding to snapping) will occur immediately, thus preventing the system from remaining in that regime.

Using Eq.~\eqref{eq:GWSolution}, we analyse the behaviour of the function $f_t(x)$ in the sub-critical and critical regimes of the BP.
It follows from the fact that the integrand in Eq.~\eqref{eq:GWSolution} is positive and diverges at $x^\prime=1$ that the value of the generating function $f_t(x)$ lies in the interval $x\le f_t(x)\le 1$ for any value of $x$ in the interval $0\le x \le 1$ and $t\ge 0$.
For such values of $x$, the value of $f_t(x)$ can be found numerically in the following way.
First, the solution, $P^*$, to the self-consistent equation~\eqref{eq:SelfConsistent} should be found.
The coefficients of the polynomial function $f(x)$ are then given by Eq.~\eqref{eq:RepxiDist}.
The integral in Eq.~\eqref{eq:GWSolution} can be evaluated in terms of the roots of the polynomial $f(x)-x$ and their residues.
Values of $f_t(x)$ can then be found by inverting Eq.~\eqref{eq:GWSolution} numerically, while derivatives of $f_t(x)$ with respect to both $x$ and $t$ can be written in closed forms in terms of $f_t(x)$.

When $f_t(x)\simeq 1$, corresponding either to $x=1$ or large values of $t$, it is possible to write $f_t(x)$ and its derivatives in terms of $P^*$ by expanding the integrand of Eq.~\eqref{eq:GWSolution} in a Taylor series around the point of divergence at $x^\prime=1$, i.e. 
\begin{eqnarray}
&&\int_{x^\prime =x}^{f_t(x)} \frac{1}{f(x^\prime)-x^\prime}\text{d}x^\prime \nonumber\\
&\simeq& \int_{x^\prime =x}^{f_t(x)} \left[\frac{1}{\left(\partial_x f(1)-1\right)\left(x^\prime-1\right)}-\frac{\partial_{xx} f(1)}{2\left(\partial_x f(1)-1\right)^2}\right]\text{d}x^\prime\nonumber\\
&\simeq&\mathbb{E}[\xi_i-1]^{-1}\left\{\ln[1-f_t(x)]-\ln[1-x]\right\}\nonumber\\
&&-2^{-1}\mathbb{E}[\xi_i(\xi_i-1)]\mathbb{E}[\xi_i-1]^{-2}\{f_t(x)-x\}~,\nonumber\\&&\label{eq:Fexpand}
\end{eqnarray}
where Eq.~\eqref{eq:RepxiDist} can be used to calculate the coefficients of the polynomial $f(x)$, giving,
\begin{eqnarray}
\partial_x f(1)-1=\mathbb{E}[\xi_i-1]&=&\left(\frac{\partial F_1}{\partial P^*}\right)-1\nonumber\\
\partial_{xx} f(1)=\mathbb{E}[\xi_i(\xi_i-1)]&=&\left(1-q^{-1}\right)\left(\frac{\partial F_0}{\partial P^*}\right)\left(\frac{\partial F_2}{\partial P^*}\right)~.\nonumber\\
\end{eqnarray}
For $x\to 1$, Eq.~\eqref{eq:Fexpand} gives for the sub-critical regime of the BP,
\begin{eqnarray}
f_t(1)&=&1\nonumber\\
\partial_xf_t(1)&=&\exp\left(\Gamma t\mathbb{E}[\xi_i-1]\right)\nonumber\\
\partial_{xx}f_t(1)&=&\frac{\mathbb{E}[\xi_i(\xi_i-1)]}{\mathbb{E}[\xi_i-1 ]}\left\{1-\exp\left(\Gamma t\mathbb{E}[\xi_i-1 ]\right)\right\}\nonumber\\&\times&\exp\left(\Gamma t \mathbb{E}[\xi_i-1 ]\right)~,\label{eq:SubCritExpand}
\end{eqnarray}
while for large $t\gg \Gamma^{-1}$,
\begin{eqnarray}
\partial_tf_t(x)&\propto&\exp\left(\Gamma t\mathbb{E}[\xi_i-1]\right)~.
\end{eqnarray}
When the BP is in the critical regime, $\mathbb{E}[ \xi_i-1]=0$ and the expansion at large $t\gg \Gamma^{-1}$ reads instead as,
\begin{eqnarray}
\partial_tf_t(x)&\simeq&\frac{2}{\mathbb{E}[\xi_i(\xi_i-1)] \Gamma t^2}~.\label{eq:critdt}
\end{eqnarray}
The value of $f_t(x)$, found numerically by solving Eq.~\eqref{eq:GWSolution} or, for certain $x$ and $t$, analytically using 
Eqs.~\eqref{eq:SubCritExpand}-\eqref{eq:critdt}, leads to the results given in Sec.~\ref{sec:TimeDep}.


\begin{thebibliography}{10}%
\makeatletter
\providecommand \@ifxundefined [1]{%
 \ifx #1\undefined \expandafter \@firstoftwo
 \else \expandafter \@secondoftwo
\fi
}%
\providecommand \@ifnum [1]{%
 \ifnum #1\expandafter \@firstoftwo
 \else \expandafter \@secondoftwo
\fi
}%
\providecommand \enquote [1]{``#1''}%
\providecommand \bibnamefont  [1]{#1}%
\providecommand \bibfnamefont [1]{#1}%
\providecommand \citenamefont [1]{#1}%
\providecommand\href[0]{\@sanitize\@href}%
\providecommand\@href[1]{\endgroup\@@startlink{#1}\endgroup\@@href}%
\providecommand\@@href[1]{#1\@@endlink}%
\providecommand \@sanitize [0]{\begingroup\catcode`\&12\catcode`\#12\relax}%
\@ifxundefined \pdfoutput {\@firstoftwo}{%
 \@ifnum{\z@=\pdfoutput}{\@firstoftwo}{\@secondoftwo}%
}{%
 \providecommand\@@startlink[1]{\leavevmode}%
 \providecommand\@@endlink[0]{}%
}{%
 \providecommand\@@startlink[1]{%
  \leavevmode
  \pdfstartlink
   attr{/Border[0 0 1 ]/H/I/C[0 1 1]}%
   user{/Subtype/Link/A<</Type/Action/S/URI/URI(#1)>>}%
  \relax
 }%
 \providecommand\@@endlink[0]{\pdfendlink}%
}%
\providecommand \url  [0]{\begingroup\@sanitize \@url }%
\providecommand \@url [1]{\endgroup\@href {#1}{\urlprefix}}%
\providecommand \urlprefix [0]{URL }%
\providecommand \Eprint[0]{\href }%
\@ifxundefined \urlstyle {%
  \providecommand \doi [1]{doi:\discretionary{}{}{}#1}%
}{%
  \providecommand \doi [0]{doi:\discretionary{}{}{}\begingroup
  \urlstyle{rm}\Url }%
}%
\providecommand \doibase [0]{http://dx.doi.org/}%
\providecommand \Doi[1]{\href{\doibase#1}}%
\providecommand \bibAnnote [3]{%
  \BibitemShut{#1}%
  \begin{quotation}\noindent
    \textsc{Key:}\ #2\\\textsc{Annotation:}\ #3%
  \end{quotation}%
}%
\providecommand \bibAnnoteFile [2]{%
  \IfFileExists{#2}{\bibAnnote {#1} {#2} {\input{#2}}}{}%
}%
\providecommand \typeout [0]{\immediate \write \m@ne }%
\providecommand \selectlanguage [0]{\@gobble}%
\providecommand \bibinfo [0]{\@secondoftwo}%
\providecommand \bibfield [0]{\@secondoftwo}%
\providecommand \translation [1]{[#1]}%
\providecommand \BibitemOpen[0]{}%
\providecommand \bibitemStop [0]{}%
\providecommand \bibitemNoStop [0]{.\EOS\space}%
\providecommand \EOS [0]{\spacefactor3000\relax}%
\providecommand \BibitemShut [1]{\csname bibitem#1\endcsname}%
\bibitem{Sethna2001}%
  \BibitemOpen
  \bibfield{author}{%
  \bibinfo {author} {\bibfnamefont{J.~P.}\ \bibnamefont{Sethna}}, \bibinfo
  {author} {\bibfnamefont{K.~A.}\ \bibnamefont{Dahmen}},\ and\ \bibinfo
  {author} {\bibfnamefont{C.~R.}\ \bibnamefont{Myers}},\ }%
  \bibfield{journal}{%
  \bibinfo {journal} {Nature (London)}\ }%
  \textbf{\bibinfo {volume} {410}},\ \bibinfo {pages} {242} (\bibinfo {year}
  {2001})%
  \bibAnnoteFile{NoStop}{Sethna2001}%
\bibitem{Durin_review2004}%
  \BibitemOpen
  \bibfield{author}{%
  \bibinfo {author} {\bibfnamefont{G.}~\bibnamefont{Durin}}\ and\ \bibinfo
  {author} {\bibfnamefont{S.}~\bibnamefont{Zapperi}},\ }%
  \enquote{\bibinfo {title} {The science of hysteresis},}\ \ (\bibinfo
  {publisher} {Elsevier},\ \bibinfo {address} {Amsterdam},\ \bibinfo {year}
  {2006})\ Chap.\ \bibinfo {chapter} {The Barkhausen effect}%
  \bibAnnoteFile{NoStop}{Durin_review2004}%
\bibitem{Pradhan2001}%
  \BibitemOpen
  \bibfield{author}{%
  \bibinfo {author} {\bibfnamefont{S.}~\bibnamefont{Pradhan}}\ and\ \bibinfo
  {author} {\bibfnamefont{B.~K.}\ \bibnamefont{Chakrabarti}},\ }%
  \bibfield{journal}{%
  \Doi{10.1103/PhysRevE.65.016113}{\bibinfo {journal} {Phys. Rev. E}}\ }%
  \textbf{\bibinfo {volume} {65}},\ \bibinfo {pages} {016113} (\bibinfo {month}
  {Dec}\ \bibinfo {year} {2001})%
  \bibAnnoteFile{NoStop}{Pradhan2001}%
\bibitem{Pradhan2009}%
  \BibitemOpen
  \bibfield{author}{%
  \bibinfo {author} {\bibfnamefont{S.}~\bibnamefont{Pradhan}}\ and\ \bibinfo
  {author} {\bibfnamefont{P.~C.}\ \bibnamefont{Hemmer}},\ }%
  \bibfield{journal}{%
  \Doi{10.1103/PhysRevE.79.041148}{\bibinfo {journal} {Phys. Rev. E}}\ }%
  \textbf{\bibinfo {volume} {79}},\ \bibinfo {pages} {041148} (\bibinfo {month}
  {Apr}\ \bibinfo {year} {2009})%
  \bibAnnoteFile{NoStop}{Pradhan2009}%
\bibitem{Carlson1989}%
  \BibitemOpen
  \bibfield{author}{%
  \bibinfo {author} {\bibfnamefont{J.~M.}\ \bibnamefont{Carlson}}\ and\
  \bibinfo {author} {\bibfnamefont{J.~S.}\ \bibnamefont{Langer}},\ }%
  \bibfield{journal}{%
  \Doi{10.1103/PhysRevA.40.6470}{\bibinfo {journal} {Phys. Rev. A}}\ }%
  \textbf{\bibinfo {volume} {40}},\ \bibinfo {pages} {6470} (\bibinfo {month}
  {Dec}\ \bibinfo {year} {1989})%
  \bibAnnoteFile{NoStop}{Carlson1989}%
\bibitem{Baro_PRL2013}%
  \BibitemOpen
  \bibfield{author}{%
  \bibinfo {author} {\bibfnamefont{J.}~\bibnamefont{Bar\'o}}, \bibinfo {author}
  {\bibfnamefont{A.}~\bibnamefont{Corral}}, \bibinfo {author}
  {\bibfnamefont{X.}~\bibnamefont{Illa}}, \bibinfo {author}
  {\bibfnamefont{A.}~\bibnamefont{Planes}}, \bibinfo {author}
  {\bibfnamefont{E.~K.~H.}\ \bibnamefont{Salje}}, \bibinfo {author}
  {\bibfnamefont{W.}~\bibnamefont{Schranz}}, \bibinfo {author}
  {\bibfnamefont{D.~E.}\ \bibnamefont{Soto-Parra}},\ and\ \bibinfo {author}
  {\bibfnamefont{E.}~\bibnamefont{Vives}},\ }%
  \bibfield{journal}{%
  \Doi{10.1103/PhysRevLett.110.088702}{\bibinfo {journal} {Phys. Rev. Lett.}}\
  }%
  \textbf{\bibinfo {volume} {110}},\ \bibinfo {pages} {088702} (\bibinfo
  {month} {Feb}\ \bibinfo {year} {2013})%
  \bibAnnoteFile{NoStop}{Baro_PRL2013}%
\bibitem{Michard-Bouchaud_EPJB2005_RFIM-OpinionDynamics}%
  \BibitemOpen
  \bibfield{author}{%
  \bibinfo {author} {\bibfnamefont{Q.}~\bibnamefont{Michard}}\ and\ \bibinfo
  {author} {\bibfnamefont{J.-P.}\ \bibnamefont{Bouchaud}},\ }%
  \bibfield{journal}{%
  \bibinfo {journal} {Eur. Phys. J. B}\ }%
  \textbf{\bibinfo {volume} {47}},\ \bibinfo {pages} {151 } (\bibinfo {year}
  {2005})%
  \bibAnnoteFile{NoStop}{Michard-Bouchaud_EPJB2005_RFIM-OpinionDynamics}%
\bibitem{Sethna1993}%
  \BibitemOpen
  \bibfield{author}{%
  \bibinfo {author} {\bibfnamefont{J.~P.}\ \bibnamefont{Sethna}}, \bibinfo
  {author} {\bibfnamefont{K.~A.}\ \bibnamefont{Dahmen}}, \bibinfo {author}
  {\bibfnamefont{S.}~\bibnamefont{Kartha}}, \bibinfo {author}
  {\bibfnamefont{J.}~\bibnamefont{Krumhansl}}, \bibinfo {author}
  {\bibfnamefont{B.}~\bibnamefont{Roberts}},\ and\ \bibinfo {author}
  {\bibfnamefont{J.}~\bibnamefont{Shore}},\ }%
  \bibfield{journal}{%
  \bibinfo {journal} {Phys.\ Rev.\ Lett.}\ }%
  \textbf{\bibinfo {volume} {70}},\ \bibinfo {pages} {3347 } (\bibinfo {year}
  {1993})%
  \bibAnnoteFile{NoStop}{Sethna1993}%
\bibitem{Jagers2005}%
  \BibitemOpen
  \bibfield{author}{%
  \bibinfo {author} {\bibfnamefont{P.}~\bibnamefont{Haccou}}, \bibinfo {author}
  {\bibfnamefont{P.}~\bibnamefont{Jagers}},\ and\ \bibinfo {author}
  {\bibfnamefont{V.~A.}\ \bibnamefont{Vatutin}},\ }%
  \emph{\bibinfo {title} {Branching Processes: Variation, Growth and Extinction
  of Populations}}\ (\bibinfo {publisher} {Cambridge University Press},\
  \bibinfo {year} {2005})%
  \bibAnnoteFile{NoStop}{Jagers2005}%
\bibitem{Sabhapandit2000}%
  \BibitemOpen
  \bibfield{author}{%
  \bibinfo {author} {\bibfnamefont{S.}~\bibnamefont{Sabhapandit}}, \bibinfo
  {author} {\bibfnamefont{P.}~\bibnamefont{Shukla}},\ and\ \bibinfo {author}
  {\bibfnamefont{D.}~\bibnamefont{Dhar}},\ }%
  \bibfield{journal}{%
  \bibinfo {journal} {J.\ Stat.\ Phys.}\ }%
  \textbf{\bibinfo {volume} {98}},\ \bibinfo {pages} {103 } (\bibinfo {year}
  {2000})%
  \bibAnnoteFile{NoStop}{Sabhapandit2000}%
\bibitem{SabhapanditPRL}%
  \BibitemOpen
  \bibfield{author}{%
  \bibinfo {author} {\bibfnamefont{S.}~\bibnamefont{Sabhapandit}}, \bibinfo
  {author} {\bibfnamefont{D.}~\bibnamefont{Dhar}},\ and\ \bibinfo {author}
  {\bibfnamefont{P.}~\bibnamefont{Shukla}},\ }%
  \bibfield{journal}{%
  \Doi{10.1103/PhysRevLett.88.197202}{\bibinfo {journal} {Phys. Rev. Lett.}}\
  }%
  \textbf{\bibinfo {volume} {88}},\ \bibinfo {pages} {197202} (\bibinfo {month}
  {Apr}\ \bibinfo {year} {2002})%
  \bibAnnoteFile{NoStop}{SabhapanditPRL}%
\bibitem{Dhar1997}%
  \BibitemOpen
  \bibfield{author}{%
  \bibinfo {author} {\bibfnamefont{D.}~\bibnamefont{Dhar}}, \bibinfo {author}
  {\bibfnamefont{P.}~\bibnamefont{Shukla}},\ and\ \bibinfo {author}
  {\bibfnamefont{J.~P.}\ \bibnamefont{Sethna}},\ }%
  \bibfield{journal}{%
  \bibinfo {journal} {J. Phys. A: Math. Gen.}\ }%
  \textbf{\bibinfo {volume} {30}},\ \bibinfo {pages} {5259} (\bibinfo {year}
  {1997})%
  \bibAnnoteFile{NoStop}{Dhar1997}%
\bibitem{Perkovic1999}%
  \BibitemOpen
  \bibfield{author}{%
  \bibinfo {author} {\bibfnamefont{O.}~\bibnamefont{Perkovi{\'c}}}, \bibinfo
  {author} {\bibfnamefont{K.~A.}\ \bibnamefont{Dahmen}},\ and\ \bibinfo
  {author} {\bibfnamefont{J.~P.}\ \bibnamefont{Sethna}},\ }%
  \bibfield{journal}{%
  \bibinfo {journal} {Phys.\ Rev.\ B}\ }%
  \textbf{\bibinfo {volume} {59}},\ \bibinfo {pages} {6106 } (\bibinfo {year}
  {1999})%
  \bibAnnoteFile{NoStop}{Perkovic1999}%
\bibitem{Spasojevi2011}%
  \BibitemOpen
  \bibfield{author}{%
  \bibinfo {author} {\bibfnamefont{D.}~\bibnamefont{Spasojevic}}, \bibinfo
  {author} {\bibfnamefont{S.}~\bibnamefont{Janic}},\ and\ \bibinfo {author}
  {\bibfnamefont{M.}~\bibnamefont{Knezevic}},\ }%
  \bibfield{journal}{%
  \Doi{10.1103/PhysRevLett.106.175701}{\bibinfo {journal} {Phys. Rev. Lett.}}\
  }%
  \textbf{\bibinfo {volume} {106}},\ \bibinfo {pages} {175701} (\bibinfo
  {month} {Apr}\ \bibinfo {year} {2011})%
  \bibAnnoteFile{NoStop}{Spasojevi2011}%
\bibitem{PerezReche2003}%
  \BibitemOpen
  \bibfield{author}{%
  \bibinfo {author} {\bibfnamefont{F.~J.}\ \bibnamefont{P\'erez-Reche}}\ and\
  \bibinfo {author} {\bibfnamefont{E.}~\bibnamefont{Vives}},\ }%
  \bibfield{journal}{%
  \bibinfo {journal} {Phys. Rev. B}\ }%
  \textbf{\bibinfo {volume} {67}},\ \bibinfo {pages} {134421} (\bibinfo {year}
  {2003})%
  \bibAnnoteFile{NoStop}{PerezReche2003}%
\bibitem{PerezReche2004RFIMField}%
  \BibitemOpen
  \bibfield{author}{%
  \bibinfo {author} {\bibfnamefont{F.~J.}\ \bibnamefont{P\'erez-Reche}}\ and\
  \bibinfo {author} {\bibfnamefont{E.}~\bibnamefont{Vives}},\ }%
  \bibfield{journal}{%
  \bibinfo {journal} {Phys. Rev. B}\ }%
  \textbf{\bibinfo {volume} {70}},\ \bibinfo {pages} {214422} (\bibinfo {year}
  {2004})%
  \bibAnnoteFile{NoStop}{PerezReche2004RFIMField}%
\bibitem{Alstrom_PRA1988}%
  \BibitemOpen
  \bibfield{author}{%
  \bibinfo {author} {\bibfnamefont{P.}~\bibnamefont{Alstr\o{}m}},\ }%
  \bibfield{journal}{%
  \Doi{10.1103/PhysRevA.38.4905}{\bibinfo {journal} {Phys. Rev. A}}\ }%
  \textbf{\bibinfo {volume} {38}},\ \bibinfo {pages} {4905} (\bibinfo {month}
  {Nov}\ \bibinfo {year} {1988})%
  \bibAnnoteFile{NoStop}{Alstrom_PRA1988}%
\bibitem{Kinouchi_PRE1999}%
  \BibitemOpen
  \bibfield{author}{%
  \bibinfo {author} {\bibfnamefont{O.}~\bibnamefont{Kinouchi}}\ and\ \bibinfo
  {author} {\bibfnamefont{C.~P.~C.}\ \bibnamefont{Prado}},\ }%
  \bibfield{journal}{%
  \Doi{10.1103/PhysRevE.59.4964}{\bibinfo {journal} {Phys. Rev. E}}\ }%
  \textbf{\bibinfo {volume} {59}},\ \bibinfo {pages} {4964} (\bibinfo {month}
  {May}\ \bibinfo {year} {1999})%
  \bibAnnoteFile{NoStop}{Kinouchi_PRE1999}%
\bibitem{Goh_PRL2003}%
  \BibitemOpen
  \bibfield{author}{%
  \bibinfo {author} {\bibfnamefont{K.-I.}\ \bibnamefont{Goh}}, \bibinfo
  {author} {\bibfnamefont{D.-S.}\ \bibnamefont{Lee}}, \bibinfo {author}
  {\bibfnamefont{B.}~\bibnamefont{Kahng}},\ and\ \bibinfo {author}
  {\bibfnamefont{D.}~\bibnamefont{Kim}},\ }%
  \bibfield{journal}{%
  \Doi{10.1103/PhysRevLett.91.148701}{\bibinfo {journal} {Phys. Rev. Lett.}}\
  }%
  \textbf{\bibinfo {volume} {91}},\ \bibinfo {pages} {148701} (\bibinfo {month}
  {Oct}\ \bibinfo {year} {2003})%
  \bibAnnoteFile{NoStop}{Goh_PRL2003}%
\bibitem{Kuntz2000}%
  \BibitemOpen
  \bibfield{author}{%
  \bibinfo {author} {\bibfnamefont{M.~C.}\ \bibnamefont{Kuntz}}\ and\ \bibinfo
  {author} {\bibfnamefont{J.~P.}\ \bibnamefont{Sethna}},\ }%
  \bibfield{journal}{%
  \bibinfo {journal} {Phys.\ Rev.\ B}\ }%
  \textbf{\bibinfo {volume} {62}},\ \bibinfo {pages} {11699 } (\bibinfo {year}
  {2000})%
  \bibAnnoteFile{NoStop}{Kuntz2000}%
\bibitem{Mehta2002}%
  \BibitemOpen
  \bibfield{author}{%
  \bibinfo {author} {\bibfnamefont{A.~P.}\ \bibnamefont{Mehta}}, \bibinfo
  {author} {\bibfnamefont{A.~C.}\ \bibnamefont{Mills}}, \bibinfo {author}
  {\bibfnamefont{K.~A.}\ \bibnamefont{Dahmen}},\ and\ \bibinfo {author}
  {\bibfnamefont{J.~P.}\ \bibnamefont{Sethna}},\ }%
  \bibfield{journal}{%
  \bibinfo {journal} {Phys.\ Rev.\ E}\ }%
  \textbf{\bibinfo {volume} {65}},\ \bibinfo {pages} {046139} (\bibinfo {year}
  {2002})%
  \bibAnnoteFile{NoStop}{Mehta2002}%
\bibitem{SpasojevicEPL2006}%
  \BibitemOpen
  \bibfield{author}{%
  \bibinfo {author} {\bibfnamefont{D.}~\bibnamefont{Spasojevic}}, \bibinfo
  {author} {\bibfnamefont{S.}~\bibnamefont{Janicevic}},\ and\ \bibinfo {author}
  {\bibfnamefont{M.}~\bibnamefont{Knezevic}},\ }%
  \bibfield{journal}{%
  \bibinfo {journal} {EPL (Europhysics Letters)}\ }%
  \textbf{\bibinfo {volume} {76}},\ \bibinfo {pages} {912} (\bibinfo {year}
  {2006})%
  \bibAnnoteFile{NoStop}{SpasojevicEPL2006}%
\bibitem{Papanikolaou_NatPhys2011}%
  \BibitemOpen
  \bibfield{author}{%
  \bibinfo {author} {\bibfnamefont{S.}~\bibnamefont{Papanikolaou}}, \bibinfo
  {author} {\bibfnamefont{F.}~\bibnamefont{Bohn}}, \bibinfo {author}
  {\bibfnamefont{R.~L.}\ \bibnamefont{Sommer}}, \bibinfo {author}
  {\bibfnamefont{G.}~\bibnamefont{Durin}}, \bibinfo {author}
  {\bibfnamefont{S.}~\bibnamefont{Zapperi}},\ and\ \bibinfo {author}
  {\bibfnamefont{J.~P.}\ \bibnamefont{Sethna}},\ }%
  \bibfield{journal}{%
  \bibinfo {journal} {Nat Phys}\ }%
  \textbf{\bibinfo {volume} {7}},\ \bibinfo {pages} {316} (\bibinfo {month}
  {Apr.}\ \bibinfo {year} {2011}),\ ISSN \bibinfo {issn} {1745-2473}%
  \bibAnnoteFile{NoStop}{Papanikolaou_NatPhys2011}%
\bibitem{Spasojevic1996}%
  \BibitemOpen
  \bibfield{author}{%
  \bibinfo {author} {\bibfnamefont{D.}~\bibnamefont{Spasojevi{\'c}}}, \bibinfo
  {author} {\bibfnamefont{S.}~\bibnamefont{Bukvi{\'c}}}, \bibinfo {author}
  {\bibfnamefont{S.}~\bibnamefont{Milo{\u{s}}evi{\'c}}},\ and\ \bibinfo
  {author} {\bibfnamefont{H.~E.}\ \bibnamefont{Stanley}},\ }%
  \bibfield{journal}{%
  \bibinfo {journal} {Phys.\ Rev.\ E}\ }%
  \textbf{\bibinfo {volume} {54}},\ \bibinfo {pages} {2531 } (\bibinfo {year}
  {1996})%
  \bibAnnoteFile{NoStop}{Spasojevic1996}%
\bibitem{Dahmen1995}%
  \BibitemOpen
  \bibfield{author}{%
  \bibinfo {author}
  {\bibfnamefont{O.}~\bibnamefont{Perkovi\ifmmode~\acute{c}\else \'{c}\fi{}}},
  \bibinfo {author} {\bibfnamefont{K.}~\bibnamefont{Dahmen}},\ and\ \bibinfo
  {author} {\bibfnamefont{J.~P.}\ \bibnamefont{Sethna}},\ }%
  \bibfield{journal}{%
  \Doi{10.1103/PhysRevLett.75.4528}{\bibinfo {journal} {Phys. Rev. Lett.}}\ }%
  \textbf{\bibinfo {volume} {75}},\ \bibinfo {pages} {4528} (\bibinfo {month}
  {Dec}\ \bibinfo {year} {1995})%
  \bibAnnoteFile{NoStop}{Dahmen1995}%
\bibitem{Durin2000}%
  \BibitemOpen
  \bibfield{author}{%
  \bibinfo {author} {\bibfnamefont{G.}~\bibnamefont{Durin}}\ and\ \bibinfo
  {author} {\bibfnamefont{S.}~\bibnamefont{Zapperi}},\ }%
  \bibfield{journal}{%
  \Doi{10.1103/PhysRevLett.84.4705}{\bibinfo {journal} {Phys. Rev. Lett.}}\ }%
  \textbf{\bibinfo {volume} {84}},\ \bibinfo {pages} {4705} (\bibinfo {month}
  {May}\ \bibinfo {year} {2000})%
  \bibAnnoteFile{NoStop}{Durin2000}%
\bibitem{Durin2001}%
  \BibitemOpen
  \bibfield{author}{%
  \bibinfo {author} {\bibfnamefont{G.}~\bibnamefont{Durin}}\ and\ \bibinfo
  {author} {\bibfnamefont{S.}~\bibnamefont{Zapperi}},\ }%
  \enquote{\bibinfo {title} {{\small{``On the power spectrum of magnetization
  noise''}}},}\ \bibinfo {howpublished} {cond-mat/0106113} (\bibinfo {year}
  {2001})%
  \bibAnnoteFile{NoStop}{Durin2001}%
\bibitem{Handford2012}%
  \BibitemOpen
  \bibfield{author}{%
  \bibinfo {author} {\bibfnamefont{T.~P.}\ \bibnamefont{Handford}}, \bibinfo
  {author} {\bibfnamefont{F.-J.}\ \bibnamefont{Perez-Reche}},\ and\ \bibinfo
  {author} {\bibfnamefont{S.~N.}\ \bibnamefont{Taraskin}},\ }%
  \bibfield{journal}{%
  \bibinfo {journal} {Journal of Statistical Mechanics: Theory and Experiment}\
  }%
  \textbf{\bibinfo {volume} {2012}},\ \bibinfo {pages} {P01001} (\bibinfo
  {year} {2012})%
  \bibAnnoteFile{NoStop}{Handford2012}%
\bibitem{Cote1991}%
  \BibitemOpen
  \bibfield{author}{%
  \bibinfo {author} {\bibfnamefont{P.~J.}\ \bibnamefont{Cote}}\ and\ \bibinfo
  {author} {\bibfnamefont{L.~V.}\ \bibnamefont{Meisel}},\ }%
  \bibfield{journal}{%
  \bibinfo {journal} {Phys.\ Rev.\ Lett.}\ }%
  \textbf{\bibinfo {volume} {67}},\ \bibinfo {pages} {1334 } (\bibinfo {year}
  {1991})%
  \bibAnnoteFile{NoStop}{Cote1991}%
\bibitem{Baldassarri2003}%
  \BibitemOpen
  \bibfield{author}{%
  \bibinfo {author} {\bibfnamefont{A.}~\bibnamefont{Baldassarri}}, \bibinfo
  {author} {\bibfnamefont{F.}~\bibnamefont{Colaiori}},\ and\ \bibinfo {author}
  {\bibfnamefont{C.}~\bibnamefont{Castellano}},\ }%
  \bibfield{journal}{%
  \bibinfo {journal} {Phys.\ Rev.\ Lett.}\ }%
  \textbf{\bibinfo {volume} {90}},\ \bibinfo {pages} {060601} (\bibinfo {year}
  {2003})%
  \bibAnnoteFile{NoStop}{Baldassarri2003}%
\bibitem{Acharyya1996}%
  \BibitemOpen
  \bibfield{author}{%
  \bibinfo {author} {\bibfnamefont{M.}~\bibnamefont{Acharyya}}\ and\ \bibinfo
  {author} {\bibfnamefont{B.~K.}\ \bibnamefont{Chakrabarti}},\ }%
  \bibfield{journal}{%
  \Doi{10.1016/0378-4371(95)00362-2}{\bibinfo {journal} {Physica A: Statistical
  Mechanics and its Applications}}\ }%
  \textbf{\bibinfo {volume} {224}},\ \bibinfo {pages} {254 } (\bibinfo {year}
  {1996}),\ ISSN \bibinfo {issn} {0378-4371}%
  \bibAnnoteFile{NoStop}{Acharyya1996}%
\bibitem{Pradhan2011}%
  \BibitemOpen
  \bibfield{author}{%
  \bibinfo {author} {\bibfnamefont{S.}~\bibnamefont{Pradhan}}\ and\ \bibinfo
  {author} {\bibfnamefont{P.~C.}\ \bibnamefont{Hemmer}},\ }%
  \bibfield{journal}{%
  \Doi{10.1103/PhysRevE.83.041116}{\bibinfo {journal} {Phys. Rev. E}}\ }%
  \textbf{\bibinfo {volume} {83}},\ \bibinfo {pages} {041116} (\bibinfo {month}
  {Apr}\ \bibinfo {year} {2011})%
  \bibAnnoteFile{NoStop}{Pradhan2011}%
\end{thebibliography}
\end{document}